\documentclass[prc,aps,showpacs]{revtex4}
\usepackage{graphicx}

\newcommand{\be}{\begin{equation}}          
\newcommand{\ee}{\end{equation}}            

\newcommand{\rptt}{\rule{0pt}{16pt}}   
\newcommand{\rpttt}{\rule{0pt}{22pt}}  
\newcommand{\dist}{\displaystyle}
\newcommand{\reac}{reaction~(\ref{1})}

\newcommand{\ega}{E_{\gamma}}

\newcommand{\bfa}{\mbox {\boldmath $a$}}       
\newcommand{\bfpi}{\mbox {\boldmath $\pi$}}       
\newcommand{\bfsig}{\mbox {\boldmath $\sigma$}} 
\newcommand{\bftau}{\mbox {\boldmath $\tau$}}     
\newcommand{\bfeps}{\mbox {\boldmath $\epsilon$}} 
\newcommand{\bfrho}{\mbox {\boldmath $\rho$}}     
\newcommand{\bfeta}{\mbox {\boldmath $\eta$}}  
\newcommand{\bfp}{\mbox {\boldmath $p$}}     
\newcommand{\bfe}{\mbox {\boldmath $e$}}     
\newcommand{\bfpp}{\mbox {\boldmath $P$}}    
\newcommand{\bfq}{\mbox {\boldmath $q$}}     
\newcommand{\bfqq}{\mbox {\boldmath $Q$}}
\newcommand{\bfk}{\mbox {\boldmath $k$}}     
\newcommand{\bfs}{\mbox {\boldmath $s$}}     
\newcommand{\bfr}{\mbox {\boldmath $r$}}     
\newcommand{\bfn}{\mbox {\boldmath $n$}}
\newcommand{\bfm}{\mbox {\boldmath $m$}}
\newcommand{\bfnz}{\mbox {\boldmath $n$}_{3\,}}
\newcommand{\bfdel}{\mbox {\boldmath $\Delta$}}  
\newcommand{\bfgamm}{\hat {\mbox {\boldmath $\Gamma$}}}   

\newcommand{\mgd}{M_{\gamma d}}           
\newcommand{\NN}{$N(1535)$}           
\newcommand{\PP}{$N(1440)$}           
\newcommand{\smunu}{\sigma_{\mu\nu}}       

\newcommand{\dij}{\delta_{ij\,}}             
\newcommand{\epsa}{\epsilon_{\mu\nu\lambda\sigma}} 

\newcommand{\dsdo}{\frac{d\sigma}{d\Omega_{_{CM}}}}  
\newcommand{\gdpid}{$\gamma d\!\to\!\pi^0d$} 
\newcommand{\gampid}{\mbox {\boldmath $\gamma d\!\to\!\pi^0d$}} 

\newcommand{\brep}{\breve\epsilon}
\newcommand{\brepa}{\breve\epsilon_1}
\newcommand{\brepb}{\breve\epsilon^{\,*}_2}
\newcommand{\abra}{\left(\!\! \begin{array}{cc} }  
\newcommand{\aket}{\end{array}\!\!\right)}         
\newcommand{\bre}{\breve e}    
\newcommand{\brq}{\breve q}    
\newcommand{\brk}{\breve k}    
\newcommand{\brs}{\breve s}    
\newcommand{\brn}{\breve n}    
\newcommand{\brm}{\breve m}    
\newcommand{\intdp}{\int\!\frac{d^3{\bf p}}{(2\pi)^3}}
\newcommand{\dbfp}{\frac{d^3{\bf p}}{(2\pi)^3}}

\newcommand{\intr}{\int\!\frac{d^3{\bf r}}{4\pi}}  
\newcommand{\intrr}{\int\!\frac{d^3{\bf r}}{(4\pi)^3}}
\newcommand{\intedr}{\int\!\frac{d^3\bfr}{(4\pi)^2}\,e^{i\bfdel\bfr}}
\newcommand{\inteqr}{\intrr\,e^{i\bfqq\bfr}}
\newcommand{\ddbfp}{\frac{d^3\bfp_2}{(2\pi)^3}\,
                    \frac{d^3\bfp'_1}{(2\pi)^3}}    

\newcommand{\intesr}{\intr e^{i\bfs\bfr}}   
\newcommand{\half}{\frac{1}{2}}
\newcommand{\intra}{\int\!\frac{d^3{\bf r}}{4\pi r}}
\newcommand{\LLaa}{\Lambda^2\!-\!\alpha^2}
\newcommand{\LLLa}{\frac{\Lambda^2_0}{\LLaa}}
\newcommand{\LLLL}{\frac{\Lambda^4_0}{\LLaa}}
\newcommand{\eLr}{e^{-\Lambda r}}
\newcommand{\ear}{e^{\alpha r}}

\newcommand{\Aij}{{\rm arctg}\frac{\Delta}{m_{ij}}}
\newcommand{\Lij}{\ln (m^2_{ij}\!+\!\Delta^2)}
\newcommand{\mL}{m_{ij}\!+\!\Lambda}
\newcommand{\scij}{\sum_{i\,j}\frac{c_{i\,}c_j}}
\newcommand{\scijp}{\scij{64\pi^2}}

\begin{document}

\title{On the reaction \gampid\ near the threshold of
       \bfeta\ production}

\author{
A.~E.~Kudryavtsev$^1$$^{,2}$\footnote[1]{Email: kudryavt@heron.itep.ru},
V.~E.~Tarasov$^1$\footnote[2]{Email: tarasov@heron.itep.ru},
I.~I.~Strakovsky$^2$\footnote[3]{Email: igor@gwu.edu},
W.~J.~Briscoe$^2$\footnote[4]{Email: briscoe@gwu.edu},
Y.~Ilieva$^2$\footnote[5]{Email: jordanka@jlab.org},
\vspace*{0.1in}
}

\affiliation{
$^1$Institute of Theoretical and Experimental Physics,
    25 Bolshaya~Cheremushkinskaya Street, Moscow 117259, Russia \\
$^2$Center for Nuclear Studies, Department of Physics,
    The George Washington University, Washington, D.C.
    20052, USA \\
}

\begin{abstract}

We consider the reaction \gdpid\ in a wide energy range around and
above the $\eta$-meson photoproduction threshold at backward CM
angles of the outgoing pion.  Our theoretical analysis is
essentially motivated by the recent measurements of the CLAS
Collaboration at Jefferson Lab, where this kinematical region of
the reaction has been thoroughly studied for the first time and a
cusps in the energy dependence of the differential cross section
in the region of $\ega\sim 600-800$~MeV has been observed. Our
preliminary and qualitative analysis, based on single- and
double-scattering diagrams, shows that the observed structure can
be explained by the contribution of the double-scattering diagram
with intermediate production of the $\eta$ meson.  The effect, to
a considerable extent, is formed due to the contribution of \NN\
resonance to the amplitudes of subprocesses on the nucleons.

\end{abstract}


\maketitle

\section{Introduction}
\label{sec:intro}

The reaction of coherent photoproduction of the $\pi^0$ meson on
the deuteron \be \gamma d\!\to\!\pi^0d \label{1}\ee has recently
been studied~\cite{CLAS} at Jefferson Lab using the CLAS detector.
The experiment was carried out in a wide range of photon energies
$\ega=0.5-2.0$~GeV at large CM scattering angles of the outgoing
pions.  A new phenomenon was observed.

The process of pion photoproduction on the nucleon $\gamma
p\!\to\!\pi N$ has been  theoretically investigated over
a long period of time (see, for example, Refs.~
\cite{Walker,Moor,Garc,Drech}).  The \reac\, on the deuteron,
has been also considered in a number of papers~\cite{Garc1,Kamal}.
During the last years, the $\eta$-photoproduction processes
have also been actively studied (see, for example, Refs.~
\cite{Knoch,Benme} for $\gamma p\!\to\!\eta p$,~\cite{Kamal,Ritz}
for $\gamma d\!\to\!\eta d$, and~\cite{Benn} for $\eta$
photoproduction on light nuclei).

The study of the meson-photoproduction processes on a deuteron
target provides information about the underlying reaction
mechanisms on few-body systems.  Our project has been motivated by
a special interest in the role of intermediate particles in \reac.

In the 1970's, the contribution of intermediate particles and
resonances to the differential cross section in the backward $\pi
d$ elastic scattering was theoretically discussed in
Ref.~\cite{Kond}.  It has been predicted that the contribution of
intermediate particles, formed in a two-step process, should
manifest itself a cusp in the energy dependence of the backward
differential cross section around the corresponding thresholds.
Such an effect, associated with intermediate $\eta$-meson
production, was confirmed by several independent measurements of
backward $\pi d$ elastic scattering~\cite{pide}.

The preliminary CLAS photoproduction data~\cite{CLAS} give for
first time clear evidence for the intermediate $\eta$-meson
effect.  A cusp in the region around the $\eta$ photoproduction
threshold, $\ega\sim 600-800$~MeV, is visible in the energy
dependence of the differential cross section at large CM
scattering angles, $\cos\theta < -0.6$. The observed effect
becomes more pronounced as the scattering angle increases. This
behavior was also seen in a previous measurement of reaction
(\ref{1})~\cite{Iman} in which a small structure was observed in
the excitation function the differential cross section at
$\cos\theta=-0.64$ (the maximum scattering angle of this
experiment). Reaction (1) was also theoretically considered
earlier (see, for example, Ref.~\cite{Garc}) in the framework of
single- and double-scattering approach without consideration of
the intermediate $\eta$ meson. A reasonable theoretical
description of the previous data~\cite{Iman} was achieved without
the necessity of inclusion of the ``$\eta$~effect"~\cite{Garc} at
low momentum transfer.

The aim of the present paper is the theoretical study of \reac\ at
large pion production angles. Our principal interest is the
contribution of the intermediate $\eta$ meson to the differential
cross section and whether it can explain the structure in the
differential cross section that has been observed in a recent CLAS
experiment~\cite{CLAS}. We shall use the standard approach based
on single- and double-scattering amplitudes.  The main
contribution to the total amplitude at large angles is expected to
come from the double-scattering terms.  We shall consider photon
energies far from the $\pi d$ threshold, say $500<\ega<1000$~MeV,
where the influence of the intermediate $N(1535)$ resonance and
the $\eta$ meson effect should be important, and it is possible to
neglect the excitation of the $\Delta(1232)$ isobar in the
intermediate state. Thus, we construct the total reaction
amplitude from terms, expected to be essential, with approximate
values of the parameters. In this paper, while we do compare our
predictions with the data qualitatively and present all the
details of our treatment, we do not attempt a detailed description
of the CLAS data; we leave this task for forthcoming papers from
the CLAS Collaboration.

This paper is organized as follows.  In Section~\ref{sec:form}, we
derive the expressions for different terms of the amplitude of
~\reac\ in a diagrammatical approach.  In Subsection
~\ref{sec:DandN}, we briefly discuss the main contributions to the
reaction amplitude and introduce the notation that we use.  In
Subsections ~\ref{sec:res} and ~\ref{sec:VME}, we give
gauge-invariant expressions for the resonance and VME
contributions to the elementary amplitudes on the nucleon.  These
results are used in Subsections~\ref{sec:single} and
~\ref{sec:double} to obtain single- and double-scattering
amplitudes of~\reac, respectively. In Section~\ref{sec:result}, we
present the numerical results. In Subsection~\ref{sec:nucl}, we
study the influence of the ``non-static" corrections that are
taken into account in the double-scattering amplitude with
intermediate $\eta$ production. In Subsection ~\ref{sec:dsg}, we
discuss our numerical results for the differential cross section
(its energy behavior at several values of $\cos\theta$) of \reac\
with backward $\pi^0$ production. The conclusion is presented in
Section~\ref{sec:conc}. In Appendices~\ref{sec:int1},
~\ref{sec:func}, ~\ref{sec:int2}, and ~\ref{sec:int3}, we
calculate the the integrals and some other expressions, used to
derive the single- and double-scattering amplitudes of \reac.

\section{Formalism}
\label{sec:form}

\subsection{Diagrams and notation}
\label{sec:DandN}

The diagrams for single- and double-scattering amplitudes
$M_{(1)}$ and $M_{(2)}$ of \reac\ are shown in Figs.~\ref{fig:g1}a
and~\ref{fig:g1}b, respectively.  The notation for the
four-momentum vectors of the initial, intermediate, and final
particles are given in this figure.  The vertices marked by
``$i\,$" or ``$j\,$" correspond to the elementary amplitudes of
the subreactions on the nucleons, and indices ``$i\,$", ``$j\,$"
specify the contributions to the elementary amplitudes considered
below.  In Fig.~\ref{fig:g1}b, the notation ``$h$" stands for the
intermediate meson.  Hereafter, we shall consider only diagrams
with $h=\pi$ or $\eta$.

The elementary photoproduction amplitude $\gamma N\!\to\!hN$ is
usually constructed as a sum of Born, vector-meson-exchange (VME),
and resonance terms~\cite{Garc,Drech,Benme}. The Born amplitudes
correspond to a set of tree diagrams with $N\!Nh$ coupling and all
possible couplings with a photon, summed by a contact $\gamma\pi
N\!N$-coupling term.  It is known that the total Born amplitude
satisfies gauge invariance (see Ref.~ \cite{Garc} and references
therein).  Using Born amplitudes for subprocesses in \reac\ on the
deuteron, one encounters the problem how to get the total
gauge-invariant amplitude. The problem comes from nucleon
off-shell effects, and the way to solve it is discussed, for
example, in Ref.~\cite{Kamal} (see also references therein).
However in our analysis, we shall neglect Born terms in the
kinematical region of reaction under study.

\begin{figure}[th]
\centering{
\includegraphics[height=0.15\textwidth, angle=0]{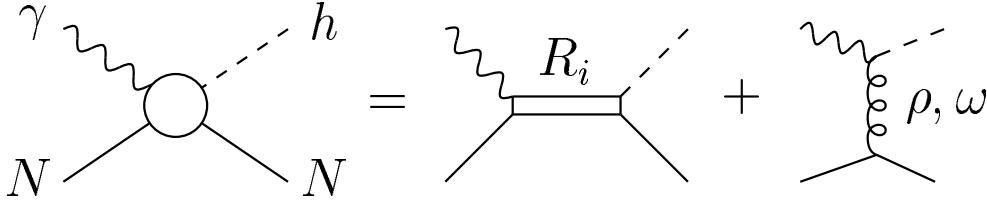}
}\caption{Diagrams of meson photoproduction on
          the nucleon (resonance and VME
          contributions).\label{fig:g2}}
\end{figure}

\begin{figure}[th]
\centering{
\includegraphics[height=0.15\textwidth, angle=0]{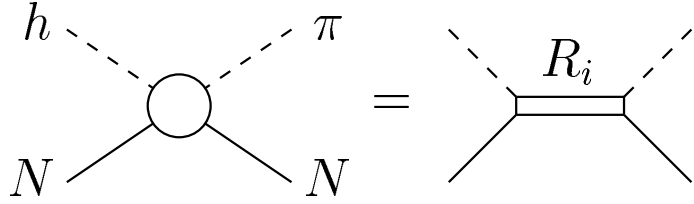}
}\caption{Diagrams of meson-nucleon binary
          reactions (resonance contributions).\label{fig:g3}}
\end{figure}

The resonance and VME terms in the photoproduction amplitude are
shown graphically in Fig.~~\ref{fig:g2}, where VME terms are
calculated via $\rho$ and $\omega$ exchanges~ \cite{Garc,Benme}.
The ``meson-meson" $hN\!\to\!\pi N$ amplitudes are written through
resonance contributions (Fig.~\ref{fig:g3}). The main contribution
from an intermediate $\eta$ meson to the cross section of \reac\
is expected from the double-scattering diagram in Fig.~
\ref{fig:g1}b ($h=\eta$), with the \NN\ excitation in both blocks
of the diagram due to the large partial width of the decay
$N(1535)\!\to\!\eta N$. Several nucleon resonances~ \cite{PDG} are
coupled to the $\pi N$ system.  Also, the couplings of \NN, \PP\,,
and $N(1520)$ to the $\eta N$ system~\cite{Garc1,Ritz,Benn} are
often used in the production amplitudes.  However in our
qualitative analysis, we shall limit the resonance parts of the
elementary amplitudes (Figs.~\ref{fig:g2} and ~\ref{fig:g3}) to
the contributions of the \NN\ in the $\pi N$ and $\eta N$ channels
and the \PP\ in the $\pi N$ channel.  We do not include the
contribution of the $\Delta(1232)$ isobar since the considered
energies are far above the $\Delta(1232)$ region.

\bigskip

Let us write the total amplitude $\mgd$ of \reac\ as

\be \mgd = M_{(1)}+M_{(2)}\, , {~~~~} M_{(1)}=\sum_i M_i\, ,
{~~~~} M_{(2)}=\sum_{i,\,h,\,j} M_{ihj}\, , \label{di1} \ee where
$h\!=\!\pi,\eta$ and index $i=1,2,\omega$, and $\rho$ (the same
for $j$) stands for \NN, \PP, $\omega$- and $\rho$-exchange for
the elementary subprocesses, respectively. Note that
$\rho$-exchange term in the single-scattering amplitude $M_{(1)}$
is forbidden by isotopic arguments.  We shall use standard
normalizations for the amplitudes, corresponding to the following
expression of the differential cross section for a binary
reaction: \be \dsdo=\frac{k}{64\pi^2 s_{\,} q}\,\overline
{|M|^2}\, . \label{di2}\ee Here, $\overline {|M|^2}$ is the square
of the total amplitude, averaged (summed) over the polarizations
of the initial (final)particles; $s$ is the square of the total CM
energy; and $q$ and $k$ are the relative 3-momenta of the initial
and final state particles.  Below, we use the following notation:
$\varphi_{1,2}$ ($\chi_{1,2}$) are Pauli spinors (isospinors) for
the initial and final nucleons in the elementary reactions or for
the nucleons in the deuteron and $\varphi^+\varphi\!=\!1$
($\chi^+\chi\!=\!1$); $\breve a\equiv ({\bf a}\bfsig)$, where
$\bfa$ is any 3-vector and $\bfsig$ is a 3-vector from the Pauli
spin matrices.

In Subsections ~\ref{sec:res} and ~\ref{sec:VME}, we give
gauge-invariant expressions for the elementary amplitudes on the
nucleon, notations for coupling constants \textit{etc.}, that will
be used for the total amplitude $\mgd$ in
Subsections~\ref{sec:single} and~\ref{sec:double}.

\subsection{Resonance terms in the amplitudes on the nucleon}
\label{sec:res}

In order to obtain the expressions for resonance contributions
to the elementary amplitudes, let us use the effective
Lagrangians for the $\pi NR$, $\,\eta NR\,$, and $\gamma NR$
interactions:
\begin{eqnarray}
L_{\pi NR}\!\!&\!\!=\!\!&\!\!-i g_{\pi NR}\,\bar N \Gamma
 \bftau R\,\bfpi +\!h.c.\, , {~~~~~}
L_{\eta NR}=-i g_{\eta NR}\, \bar N \Gamma R\, \eta +\!h.c.\, ,
 \label{a1} \\
\rptt
L_{\gamma NR}\!\!&\!\!=\!\!&\!\!\frac{e}{2(m+m_{_R})}\,
 \bar R (k^s_{_R}+k^v_{_R}\tau_3)\Gamma_{\mu\nu} N F^{\mu\nu} + h.c.
 {~~~~~} (e^2/4\pi\approx 1/137),
\label{a2} \\
\rpttt
\Gamma\!\!\!&=&\!\!1,{~~~~~~}
\Gamma_{\mu\nu}=\gamma_5\smunu {~~~~~~}({\rm odd~parity}),
 \label{a3} \\
\Gamma\!\!\!&=&\!\!\gamma_5\, ,{~~~~~}
\Gamma_{\mu\nu}=\smunu {~~~~~~~~}({\rm even~parity})
 \label{a4}
\end{eqnarray}
(we use the pseudoscalar couplings in Eq.~(\ref{a1})).  Here
${\dist \smunu\!=\!\frac{1}{2i}
(\gamma_{\mu}\gamma_{\nu}\!-\!\gamma_{\nu} \gamma_{\mu})}\,$;
$F^{\mu\nu}\!=\!\partial^{\nu}\!A^{\mu}\!-\!
\partial^{\mu}\!A^{\nu}\,$; $N,\, \bfpi,\, \eta,\, A^{\mu}$, and
$R$ are the nucleon, $\eta$, $\pi$, photon, and
resonance-particle ($R$) fields; $m$ and $m_{_R}$ are the
nucleon and resonance masses; $k^s_{_R}\,$ and $k^v_{_R}\,$
correspond to isoscalar and isovector $\gamma NR$ couplings.
Operator structures in Eqs.~(\ref{a3}) and (\ref{a4}) correspond
to odd- [\NN, $I(J^{PC})=\frac{1}{2}(\frac{1}{2}^-)$] and even-
[\PP, $I(J^{PC})=\frac{1}{2}(\frac{1}{2}^+)$] parity resonances
$R$.

Hereafter, we use the photon couplings $\hat g_{\gamma\,i\,}$,
$g_{s\,i\,}$, and $g_{v\,i\,}$, defined as
\be
\hat g_{\gamma\,i}=g_{s\,i}+\!g_{v\,i} \tau_3=
\frac{e}{m\!+\!m_i}\,(k^s_{_R}+\!k^v_{_R}\tau_3)\, ,
\label{a5}\ee
where $i = 1[N(1535)], 2[N(1440)]$ specifies resonances and their
masses $m_i$.

Below, we use the non-relativistic deuteron wave function (DWF) to
obtain the amplitudes of \reac\ (Subsections~\ref{sec:single}
and~\ref{sec:double}).  In this connection, we derive the
elementary amplitudes in a non-relativistic approximation before
they are used to obtain the amplitudes on the deuteron, leaving
the leading terms only with respect to the relative momenta.  The
relative accuracy of this approximation near the $\eta$
photoproduction threshold ($\sqrt{s}\sim m\!+\!m_{\eta}$) is of
the order of $q/2m\sim 0.2-0.3$ ($q$ is the photon relative
momentum).  Let us introduce some useful notation for the
resonance amplitude $A_{aib}$ of the reaction $a
N\!\!\to\!R_i\!\!\to\!bN$ as \be A_{aib}=\chi^+_2 \hat T \chi_1\,
\varphi^+_2 \hat S \varphi_1\, , {~~~} \hat T=\hat T_{aib}\, ,
{~~~} \hat S=\hat S_{aib}\, , \label{a6}\ee where $\hat S$ and
$\hat T$ are spin and isospin operators. Let us first consider the
spin operators, $A_{\gamma ih}$, of the photoproduction
amplitudes.  In our approximation, they have the following
structures:
$$
\hat S_{\gamma 1h}\sim (q_{0\,}\bre -e_{0\,}\brq)\, , {~~~~}
\hat S_{\gamma 2h}\sim \brk\,(\brq\,\bre -q_{0\,} e_0)\, ,
$$
where $\bre=(\bfsig\bfe)$, $\brq=(\bfsig\bfq)$, and
$\brk=(\bfsig\bfk)$; $\bfk$ ($\bfq$) is the CM 3-momentum of
the final meson (initial photon); $q_0$ and $e=(e_0,\bfe)$
are CM energy and polarization 4-vector of the photon.  Note
that gauge invariance, guaranteed by the Lagrangian $L_{\gamma
NR}$~(\ref{a2}), is obviously valid for the operators $\hat
S_{\gamma 1h}$ and $\hat S_{\gamma 2h}$.  Hereafter, we shall
fix the photon 4-vector by the gauge condition
$e_0\!=\!(\bfe\bfq)\!=\!0$.  Finally, we obtain the following
expessions for the resonance amplitudes $A_{aib}$ in terms of
the operators $\hat T_{aib}$ and $\hat S_{aib}$~(\ref{a6}):
\be
\hat T_{\gamma i\pi}=(\bfpi \bftau) \hat g_{\gamma i}\, , {~~}
\hat T_{\gamma i\eta}=\hat g_{\gamma i}\, , {~~}
\hat T_{\pi_1 i\pi_2}=(\bfpi_2 \bftau) (\bfpi_1 \bftau)\, ,
{~~} \hat T_{\eta i\pi}=(\bfpi \bftau) {~~}(i\!=\!1,2)\, ,
\label{a7}\ee
$$
\hat S_{\gamma 1 h}=iC_{\gamma 1 h\,} F_{1h}\,\bre\, ,{~~}
\hat S_{\gamma 2 h}=iC_{\gamma 2 h\,} F_{2h}\,
\brk_{\,}\brq_{\,}\bre\, , {~~}
\hat S_{h 1 \pi}=iC_{h 1 \pi\,} F_{1h\,} F_{1\pi} \,{\bf I}
{~~} (h\!=\!\pi,\eta)\, ,
$$ $$
C_{\gamma 1 h}=g_{1h}\,BW_1\,2mq_0\, , {~~~~}
C_{\gamma 2 h}=g_{2h}\,BW_2\, , {~~~~}
C_{h 1\pi}=ig_{1h\,}g_{1\pi}\,BW_1\, 2m\, ,
$$ $$
\hat S_{h 2 \pi}=iC_{h 2 \pi\,} F_{2h\,} F_{2\pi}\,\brk_{\,}\brq\, ,
{~~} C_{h 2\pi}=\frac{i}{2m}\,g_{2h\,}g_{2\pi}\,BW_1\, , {~~}
BW_i=\frac{2m_i}{s\!-\!m^2_i+i\sqrt{s}\,\Gamma_i(s)}\, ,
$$
where, in addition to the above-given notations, $\bfq$ is the CM
3-momentum of the initial photon or meson; $BW_i\,$,
$\Gamma_i(s)\,$, and $g_{ih}$ are the Breit-Wigner propagator,
total width, and coupling constant to the $hN$ channel for $i$-th
resonance, respectively; \textbf{I} is the unit $2\!\times\! 2$
matrix; and $F_{ih}$ are the form factors of the strong decays
$R_i\!\to\!hN$.  Here, we use the form factor only for $p$-wave
$R_{11}N\pi$ vertex ($F_{ih}\equiv 1$ for other vertices) in order
to compensate its energy growing in the region far away from the
$\pi N$ threshold. We take the function $F_{2\pi}=F$ in monopole
parametrization (see Eqs.~(\ref{a8})) which is convenient for
analytical calculations of the integrals in
Subsection~\ref{sec:single}.
The widths $\Gamma_i\,$, coupling constants $g_{ih}\,$, and
relative 3-momenta $q_{\,h}$ ($h=\pi,\eta$)  for the decays
$N(1535)\!\to\!\pi N,\,\eta N\,$ and $N(1440)\!\to\!\pi N$
are connected by the relations
\be
\Gamma_1(s)=\Gamma_{1\pi}(s)\!+\!\Gamma_{1\eta}(s)\, , {~~~~~~}
\Gamma_{1\pi}=3\, g^2_{1\pi}\,
\frac{(E\!+\!m)\, q_{\pi}}{4\pi\,\sqrt{s}}, {~~~~~~}
\Gamma_{1\eta}= g^2_{1\eta}\,
\frac{(E\!+\!m)\, q_{\eta}}{4\pi\,\sqrt{s}},
\label{a8}\ee
$$
\Gamma_{2\pi}=3\, g^2_{2\pi}\,\frac{(E\!-\!m)\,q_{\pi}}
{4\pi\,\sqrt{s}}\,F^2(q_{\pi}), {~~~~} F(q_{\pi})
=\frac{\Lambda^2_0}{\Lambda^2\!+\!q^2_{\pi}}\, , {~~~}
\Lambda^2_0=\Lambda^2\!+\!q^2_{0\pi}\, ,
$$
where $E\!+\!m\approx 2m$, $\,E\!-\!m\approx q^2_{\pi}/(2m)$
($E$ is nucleon total energy), and $q_{0\pi}$ is the relative
momentum in the decay $N(1440)\!\to\!\pi N$ at resonance mass.

The photon couplings $\hat g_{\gamma i}$ can be expressed
through helicity amplitudes $A^p_{1/2}$ and $A^n_{1/2}$
~\cite{Benme} of the decays $R_i\!\to\!p\gamma$ and
$R_i\!\to\!n\gamma$, respectively.  We can relate the
radiative widths to the amplitudes $A^{p,n}_{1/2}$ (for
spin-1/2 resonances) as well as to the constants $g^{p,n}
_{\gamma i}=g_{si}\pm g_{vi}$, \textit{i.e.},
\be
\Gamma(R_i\!\to\!p\gamma,\,n\gamma)=\frac{k^2_{\gamma}\,m}{\pi\,m_i}\,
|A^{p,n}_{1/2}\,|^2\, =\,\left(g^{p,n}_{\gamma i}\right)^2\,
\frac{k^3_{\gamma}}{\pi}\, ,
\label{a9}
\ee
%
where $k_{\gamma}$ is the relative photon 3-momentum in the decay.
Then, we obtain \be |A^{p,n}_{1/2}\,|^2\, =\,\left(g^{p,n}_{\gamma
i}\right)^2\, \frac{m_i}{m}\,k_{\gamma}\, , {~~~~~~}
2\,g_{vi}=(A^p_{1/2}-A^n_{1/2})\,\sqrt{\frac{m}{m_i\,k_{\gamma}}}\,
. \label{a10} \ee Note that only the isovector constants $g_{vi}$
(not $g_{si}$) are needed to derive the amplitude for \reac\ on
the deuteron.  The helicity amplitudes are usually extracted from
the photoproduction experiments, and the values
$\,A^{p,n}_{1/2}\,$ for \NN, \PP\, and other nucleon resonances
can be found, for example, in Refs. ~\cite{Drech,sm02}.

\subsection{Vector-meson exchange (VME) terms}
\label{sec:VME}

In order to derive $\omega$- and $\rho$-exchange amplitudes
of the photoreactions on the nucleon, we use the effective
Lagrangians $L_{_{VNN}}$ and $L_{em}$ of $V\!N\!N$($V=
\omega,\rho$)- and $Vh\gamma\,$($h\!=\!\pi,\eta$)-interaction,
taken in forms used in Refs.~\cite{Garc,Benme}
\be
L_{_{VNN}}=-g_{_{VNN}}\,\bar N
\left[\left(\gamma_{\mu}+\frac{\beta_{_V}}{2m}\sigma_{\mu\nu}
\partial^{\nu}\right)\left(\omega^{\mu}+\bftau\bfrho^{\mu}\right)
\right] N\, ,
\label{b1}\ee
\be
L_{em}=\epsa\,(\partial^{\mu}e^{\nu})
\left[ \frac{G_{V\pi\gamma}}{m_{\pi}}\,(\partial^{\lambda}\pi_i)
(\delta_{i3}\,\omega^{\sigma}+\rho^{\sigma}_i)] +
\frac{G_{V\eta\gamma}}{m_{\eta}}\,(\partial^{\lambda}\eta)\,
\delta_{i3}\,\rho^{\sigma}_{i\,}\right]\, \label{b2}\ee
($\epsilon_{1230}=1\,$, $\epsilon_{123}=1$), where $\mu, \nu,
\lambda, \sigma\,$ and $i, j\,$ are Lorentz and isotopic indices,
respectively, and $\pi,\,\eta,\,\rho$, and $\omega$ stand for
$\pi,\,\eta,\,\rho$, and $\,\omega$ mesons. The coupling constants
$G_{Vh\gamma\,}(h=\pi,\eta)$ in Eq. ~(\ref{b2}) can be expressed
through the radiative widths $\Gamma_{V\!\to h\gamma}$ by the
following relation: \be \Gamma_{V\!\to h\gamma}\, =
\frac{G^2_{Vh\gamma}\,q^3_{h\gamma}}{12\pi m^2_h}\,=
\frac{G^2_{Vh\gamma}}{12\pi m^2_h}\,\frac{m^3_{_V}}{8}\,
\left(1-\frac{m^2_h}{m^2_{_V}}\right)^3\, . \label{b3}\ee Using
Eqs.~(\ref{b1}) and (\ref{b2}), one can write the VME amplitudes
of the reactions $\gamma N\!\to\!h N$ as $A_{\gamma
Vh}=\chi^+_2\hat T\chi_1\,\bar u_2\hat M_{\gamma Vh}\,u_1\,$,
where $u_{1,2}$ are nucleon Dirac spinors ($\bar
u_{1,2}u_{1,2}\!=\!2m$), and \be \hat M_{\gamma
Vh}=\frac{G_{Vh\gamma}}{m_h}\,
\frac{g_{_{VNN}}}{(r^2\!-\!m^2_{_V})}\,\epsa\,q^{\mu} e^{\nu}
k^{\lambda} \left[-(1+\beta_{_V})\gamma^{\sigma}+
\frac{\beta_{_V}}{m}\,p^{\sigma}_1\,\right]\, , \label{b4}\ee
where $q^{\mu}$ and $e^{\nu}$ are the 4-momentum and polarization
of the photon, $k^{\lambda}$ and $p^{\sigma}$ are 4-momenta of the
final meson $h$ and the initial nucleon, and $r^2=(q\!-\!k)^2$ is
the 4-momentum transfer to the deuteron squared. $\hat T\!=\!\hat
T_{\gamma Vh}$ is the isospin operator: \be
\hat T_{\gamma\rho\pi}=(\bfpi\bftau)\, , {~~~}
\hat T_{\gamma\omega\pi}=(\bfnz \bfpi)\,{\bf I} , {~~~}
\hat T_{\gamma\rho\eta}=\tau_3\, ,
\label{b5}\ee
where $\bfnz\!=\!(0,0,1)$ is the unit vector in isotopic
space.
Using the notation~(\ref{a6}) for the amplitude $A_{\gamma Vh}$,
\textit{i.e.}, $\bar u_2 \hat M_{\gamma Vh}\, u_1= \varphi^+_2
\hat S_{\gamma Vh}\,\varphi_1\,$, and the gauge invariance
condition $e_0=(\bfe\bfq)=0$, in non-relativistic approximation,
we have \be \hat S_{\gamma Vh}=iC_{\gamma Vh}\, {\rm
Tr}\{\brk\,\brq\,\bre\}\,{\bf I}\, , {~~} C_{\gamma
Vh}=\frac{m}{m_h}\, \frac{G_{Vh\gamma\,}
g_{_{VNN}}}{(r^2\!-\!m^2_{_V})}\, , {~~} {\rm Tr}\{\brk
\brq\,\bre\}=2i\,(\bfk\cdot[\bfq\!\times\!\bfe])\, \label{b6}\ee
(here, the expression for the trace Tr$\{\brk \brq\,\bre\}$ is
used for convenience in the next Subsection~\ref{sec:single}). The
amplitudes (\ref{b6}) in our approximation do not contain the
constants $\beta_{_V}$ from Eq.~(\ref{b1}).  The gauge invariance
of the amplitude~(\ref{b4}) is obvious and is also satisfied in
the expression~(\ref{b6}) for $\hat S_{\gamma Vh}$ due to the
gauge-invariant factor ${\rm Tr}\{\brk \brq\,\bre\}$.

Note that the $\rho\pi\gamma$ vertex in the Lagrangian~(\ref{b2})
corresponds to isoscalar photon coupling, while only isovector
photon coupling can contribute to the amplitude of \reac.
Generally, the isotopic $\rho\pi\gamma$ vertex has the structure
$g_{1\,}(\bfpi\bfrho)+g_{2\,}\pi_3\rho _3+g_{3\,} (\bfnz
[\bfpi\!\times\!\bfrho])$, where $g_3$ is an isovector coupling
constant.  Then, for radiative $\rho$-decays, we have
$\Gamma_{\rho^0\to\pi^0\gamma}\!\sim\!(g_1\!+\!g_2)^2$ and
$\Gamma_{\rho^{\pm}\to\pi^{\pm}\gamma}\!\sim g^2_1\!+\!g^2_3$.
From the PDG~\cite{PDG}, $\Gamma_{\rho^0\to\pi^0\gamma}/
\Gamma_{\rho^{\pm}\to\pi^{\pm}\gamma}\sim 1.7\,$-$\,1.8$,
\textit{i.e.}, $(g_1\!+\!g_2)^2 > g^2_1\!+\!g^2_3$. Since
isoscalar $\rho\pi\gamma$ coupling ($g_2\!=\!g_3\!=\!0$) is
successfully used in the $\rho$-exchange amplitude of the reaction
$\gamma N\!\to\!\pi N$, we may suppose that $g_{2,3}\ll g_1$.  In
addition, let us compare $\omega$- and $\rho$-exchange amplitudes
$M_{\gamma\omega\pi}$ and $M_{\gamma\rho\pi}$ of the reaction
$\gamma N\!\to\!\pi^0N$.  Using coupling constants
from~\cite{Garc} ($g_1\equiv G_{\rho\pi\gamma}$), we obtain
$M_{\gamma\omega\pi}/M_{\gamma\rho\pi}\sim G_{\omega\pi\gamma\,}
g_{\omega NN} / G_{\rho\pi\gamma\,} g_{\rho NN}\sim 10$. Based on
that, we neglect the $\rho$-exchange amplitude with intermediate
pion production in \reac\ in comparison with $\omega$-exchange
amplitude.

\subsection{Single-scattering amplitude of the reaction
$\gamma d\to\pi^0d$}
\label{sec:single}

Let us write the amplitude $A$ of the process $\gamma
N\!\to\!\pi^0N$ in the form~(\ref{a6}), where $\hat S$ and $\hat
T$ are the spin and isospin parts of the transition operator.
Then, a single-scattering amplitude $M_{(1)}$ for \reac\ reads \be
iM_{(1)}= {\rm Tr}\{\hat T\}\int\!\frac{d\varepsilon}{2\pi}\,\dbfp
\,{\rm Tr}\{i(\bfeps^{\,*}_2\,\bfgamm^{}_2)\,iG(p_2)\,i\hat S\,
iG(p_1)\,i(\bfeps_{1\,}\bfgamm_1)\,iG(p)\}\, \label{ss1}\ee (we
follow the diagramatical technique of Ref.~\cite{Tar}, and some
comments will be given in Subsection ~\ref{sec:double}). Here:
$G(p_{1,2})=(2m\varepsilon_{1,2}\!-\!\bfp^2_{1,2}\!+\!i0)^{-1}$
and $G(p)=(2m\varepsilon\!-\!\bfp^2\!+\!i0)^{-1}$ are
nonrelativistic propagators of the intermediate nucleons in
Fig.~\ref{fig:g1}a with 3-momenta (kinetic energies)
$\bfp_{1,2\,}$ and $\bfp$ ($\varepsilon_{1,2}$ and
$\varepsilon\,$), respectively; $\bfeps_{1,2}$ and $\bfgamm_{1,2}$
are 3-vectors of polarization and $dNN$-vertex operators for
initial and final deuterons.  The vertex $\bfgamm$ is connected
with the DWFs $\Psi$ by the relations \be
(\bfeps\bfgamm)=4\sqrt{m}\,(\bfq^2\!+\!m\varepsilon_d)\hat \Psi\,
, {~~~} \Psi=\varphi^+_2\hat \Psi\,\sigma_2\varphi^*_1\,
\chi^+_2\frac{\tau_2}{\sqrt{2}}\chi^*_1\, , \label{ss2}\ee
$$
\hat \Psi=\hat \Psi(\bfeps,\bfq)=
\frac{u(q)}{\sqrt{2}}\,\brep-\frac{w(q)}{2}\left(
\frac{3\,(\bfq\bfeps)}{q^2}\brq-\brep\right),
$$
where $\bfq$ is the relative 3-momentum of nucleons,
$\varepsilon_d$ is the deuteron binding energy, $u(q)$ and $w(q)$
are the $s$- and $d$-wave parts of DWF normalized as
$\int\!d^3\bfq\,[u^2(q)\!+\!w^2(q)]\!=\!(2\pi)^3$.

Integrating Eq.~(\ref{ss1}) over the energy $\varepsilon$ and
using Eqs.~(\ref{ss2}), we obtain
\be
M_{(1)}= 2\,{\rm Tr}\{\hat T\}\int\!\dbfp\,
{\rm Tr}\{\hat\Psi^+_2\,\hat S\,\hat\Psi_1\}\, , {~~~}
\hat \Psi_{1,2} =\hat \Psi(\bfeps_{1,2\,},\bfq_{1,2})\, ,
\label{ss3}\ee
where $\bfq_{1,2}=\bfp\!-\!\half\bfpp_{1,2}$ (see
Fig.~\ref{fig:g1}a).  Then, using Eqs.~(\ref{a6}), (\ref{a7}),
(\ref{b5}), and (\ref{b6}) for the $\gamma N\!\to\!\pi^0N$
amplitudes, we obtain
\be
M_{(1)}=M_1+M_2+M_{\omega}\, , {~~~}
M_i=x_i\int\!\dbfp\,{\rm Tr}\{\hat\Psi^+_2\,\hat O_i\,\hat\Psi_1\}
{~~~}(i=1,2,\omega)\, ,
\label{ss4}\ee
$$
\hat O_1=\bre\, , {~~~~~}
\hat O_2=\brk\,\brq\,\bre\, , {~~~~~}
\hat O_{\omega}={\rm Tr}\{\brk\,\brq\,\bre\}\,{\bf I}\, ,
$$
$$
x_1=4i\,g_{v1\,} C_{\gamma 1\pi}\, , {~~~}
x_2=4i\,g_{v2\,} C_{\gamma 2\pi}\, F(k) , {~~~}
x_{\omega}=4i\, C_{\gamma \omega \pi}\, .
$$
Hereafter, $q_0\,$, $\bfq$, and $\bfk$ are the CM photon energy
and the CM 3-momenta of the initial photon and final pion,
respectively.  The values $x_{1,2}\,$, factored out of the
integrals in Eq.~(\ref{ss4}), depend on the effective mass $m_{\pi
N}\!=\!\sqrt{s}$ in the subprocess $\gamma N\!\to\!\pi N$, and we
calculate the value $m_{\pi N}$ using the 3-momentum
$\bfp\!=\!\frac{1}{4}(\bfpp_1\!+\!\bfpp_2)$ of the intermediate
nucleon in Fig.~\ref{fig:g1}a.
Expressing DWF, given by Eq.~(\ref{ss2}) for $\hat\Psi$ in
$r$-representation, where
\be
\hat\Psi(\bfeps,\bfq)=\intr e^{-i\bfq\bfr}\hat\Phi(\bfeps,\bfr)\, ,{~~~}
\hat\Phi(\bfeps,\bfr)=\frac{u(r)}{r\sqrt{2}}\,\brep-\frac{w(r)}{2r}
\left(\frac{3\,(\bfr\bfeps)}{r^2}{\breve r}-\brep\right),
\label{ss5}\ee
we obtain the amlitudes $M_i$~(\ref{ss4}) in the form
\be
M_i=x_i\intedr\,{\rm Tr}\{\hat\Phi^+_2\,\hat O_i\,\hat\Phi_1\}\, ,
{~~~}\hat\Phi_{1,2}=\hat\Phi(\bfeps,\bfr)\, , {~~~}
\bfdel=\half (\bfk\!-\!\bfq)\, .
\label{ss6}\ee

In order to evaluate the amplitudes $M_i$, let us introduce
the integrals
\be
\intedr f^2_1(r)=A_1\, ,{~~~} \intedr f_{1,2}(r)f_2(r)
\frac{r_i r_j}{r^2}= n_i n_j B_{1,2}+\dij C_{1,2}\, ,
\label{ss7}\ee
where
\be
f_1(r)=\frac{u(r)}{r\sqrt{2}}+\frac{w(r)}{2r}\, , {~~~~~}
f_2(r)=\frac{3w(r)}{2r}\, , {~~~~~} {\bf n}=\frac{\bfdel}{\Delta}
{~~}(|{\bf n}|\!=\!1).
\label{ss8}\ee
From Eqs.~(\ref{ss4})--(\ref{ss8}), we obtain the amplitudes
$M_i$ ($i=1,2,\omega$) in the form
\begin{eqnarray}
\rptt M_1 \!\!&=&\!\! x_{1\,} (A_1\!-\!2C_1)\,
 {\rm Tr}\{\brepb\,\bre\,\brepa\}
 + x_1 B_1\,{\rm Tr}\{\breve V\,\bre\,\brn\}\, ,
\label{ss9} \\
\nonumber
\rptt M_2 \!\!&=&\!\! x_{2\,} (A_1\!-\!2C_1)\,
  {\rm Tr}\{\brepb\brk\,\brq\,\bre\,\brepa\}
  + x_2 B_1\,{\rm Tr}\{\breve V \brk\,\brq\,\bre\,\brn\} \\
\nonumber
\rptt  &+&\!\! x_2 \left[ C_2\,(\bfeps_{1\,}\bfeps^*_2) +
      (B_2\!-\!2B_1)\,(\bfn\bfeps_1)(\bfn\bfeps^*_2)\right]
      {\rm Tr}\{\brk\,\brq\,\bre\}\, , \\
\nonumber
\rptt M_{\omega} \!\!&=&\!\! 2x_{\omega}\,
 \left[(A_1\!-\!2C_1\!+\!C_2)\,(\bfeps_{1\,}\bfeps^*_2)
 +(B_2\!-\!2B_1)\,(\bfn\bfeps_1)(\bfn\bfeps^*_2)\,\right]
 {\rm Tr}\{\brk\,\brq\,\bre\}\, ,
\end{eqnarray}
where
$\breve V\!=\!(\bfn\bfeps^*_2)\,\brepa\!-\!(\bfn\bfeps_1)\,\brepb\,$
($\brepb\!=\!\bfeps^{\,*}_2\bfsig\,$).
Neglecting the $d$-wave component of DWF, \textit{i.e.}, setting
$w(r)\!=\!0$ in Eq.~(\ref{ss8}), one obtains
$B_{1,2}\!=\!C_{1,2}\!=\!0$ and it simplifies Eqs.~(\ref{ss9}).
However in a single-scattering amplitude, the momentum is
transferred to one nucleon, and at large angles of the outgoing
$\pi^0$ the relative momenta $q_{1,2}$ of the nucleons become
large and the $d$-wave part of DWF should be important.  We use a
parametrization of DWF employing the Bonn potential~\cite{Bonn}
(full model) and the corresponding analytical expressions for
$A_1\,$, $B_{1,2}\,$, and $C_{1,2}\,$, used in Eqs.~(\ref{ss9}),
are given in Appendix~\ref{sec:int1}.

\subsection{Double-scattering amplitude of the reaction
$\gamma d\to\pi^0d$}
\label{sec:double}

Let $\hat S_1$($\hat S_2$) and $\hat T_1$($\hat T_2$) be spin and
isospin operators in the amplitude~(\ref{a6}) of the subprocess
$\gamma N\!\to\!h N$ ($h N\!\to\!\pi^0N$) in the diagram of
Fig.~\ref{fig:g1}b.  Then the double-scattering amplitude
$M_{(2)}$ has the form
\begin{eqnarray}
\nonumber
iM_{(2)} \!\!&=&\!\! {\rm Tr}\{\hat T_1\,\hat T^c_2\}\int\!
\frac{d\varepsilon_2}{2\pi}\,\frac{d\varepsilon'_1}{2\pi}\,\ddbfp\, \\
 \!\!&\times &\!\! {\rm Tr}\{i(\bfeps^{\,*}_2\,\bfgamm^{}_2)\,iG(p'_1)\,
 i\hat S_1\,iG(p_1)\,i(\bfeps_{1\,}\bfgamm_1)\,iG(p_2)\,
 i\hat S^c_2\,iG(p'_2)\}\,iG_h(s)\, .
\label{dd1}
\end{eqnarray}
Here, $G(p_{1,2})$ and $G(p'_{1,2})$ are nonrelativistic
propagators of the intermediate nucleons with 3-momenta (kinetic
energies) $\bfp_{1,2}$ and $\bfp'_{1,2}$ ($\varepsilon_{1,2}$ and
$\varepsilon'_{1,2}$); $G_h(s)$ is the propagator of the
intermediate meson $h$ with 3-momentum $\bfs$; $\hat T^c_2=\tau_2
\hat T^T \tau_2\,$, $\hat S^c_2=\sigma_2 \hat S^T
\sigma_2$~\cite{Tar1}, where index ``$T$" stands for transposition
operator. Integrating over the energies $\varepsilon_2$ and
$\varepsilon'_1$, and using Eq.~(\ref{ss2}), we obtain
\be M_{(2)} =-\frac{1}{m}\,{\rm Tr}\{\hat T_1\,\hat T^c_2\}\int\!
\ddbfp\,{\rm Tr}\{\hat\Psi^+_2\,\hat S_1\,\hat\Psi_1\, \hat
S^c_2\}\,G_h(s)\, , \label{dd2} \ee Inserting $\hat T$ and $\hat
S$ from Eqs.~ (\ref{a7}), (\ref{b5}), and (\ref{b6}), we obtain
the contributions $M_{ihj}$ to $M_{(2)}$ in the form: \be M_{ihj}
=-y_{ihj}\int\!\ddbfp\,{\rm Tr}\{\hat O_{ihj}\}\,G_h(s)\, ,
\label{dd3} \ee
$$
y_{\,ih1}=\frac{2g_{vi}}{m}\,C_{\gamma ih\,} C_{h1\pi}\, , {~~~}
\hat O_{1h1}=\hat\Psi^+_{2\,}\bre\,\hat\Psi_1\, , {~~~}
\hat O_{2\pi 1}=\hat\Psi^+_{2\,} \brs\,\brq\,\bre\,\hat\Psi_{1\,}
 \brs\,\brk\,F(s)\, ,
$$
$$
y_{\,ih2}=\frac{2g_{vi}}{m}\,C_{\gamma ih\,} C_{h2\pi\,} F(k)\, ,{~~}
\hat O_{1\pi 2}=\hat\Psi^+_{2\,}\bre\,\hat\Psi_1\brs\,\brk\,F(s)\, ,
{~~} \hat O_{2\pi 2}=\hat\Psi^+_{2\,}\brs\,\brq\,\bre\,\hat\Psi_{1\,}
\brs\,\brk\, F^2(s)\, ,
$$
$$
y_{\,\omega\pi 1}=-\frac{2}{m}\,C_{\gamma\omega\pi\,}
 C_{\pi 1\pi\,}\, ,{~~~} y_{\,\rho\eta 1}
 =\frac{2}{m}\,C_{\gamma\rho\eta\,} C_{\eta 1\pi\,}\, ,{~~~}
\hat O_{\omega\pi 1}=\hat O_{\rho\eta 1}=
\hat\Psi^+_{2\,}\hat\Psi_1\,{\rm Tr}\{\brs\,\brq\,\bre\}\, ,
$$
$$
y_{\,\omega\pi 2}=-\frac{2}{m}\,C_{\gamma\omega\pi\,}
 C_{\pi 2\pi\,} F(k)\, ,{~~~~} \hat O_{\omega\pi 2}=\hat\Psi^+_{2\,}
\hat\Psi_{1\,}\brs\,\brk\,F(s)\,{\rm Tr}\{\brs\,\brq\,\bre\}\, ,
$$
where $F(s)$ is a form factor in the $N(1440)N\pi$ vertex
(see Eq.~(\ref{a8})).

In the case of the double-scattering amplitude due to DWF, the
main contribution to the integral $\int\!d^3\bfp_2 d^3\bfp'_1$
comes from the regions $\bfp_{1,2}\sim \half\bfpp_1$ and
$\bfp'_{1,2}\sim \half\bfpp_2$ with small relative momenta
$q_{1,2}\sim 0$. To simplify the calculations, we neglect the
$d$-wave components of DWF in the amplitudes $M_{ihj}$. The
factors $y_{ihj}$ in Eqs.~(\ref{dd3}) are calculated at \be
\bfp_1\!=\!\bfp_2\!=\!\half\bfpp_1\, , {~~~~}
\bfp'_1\!=\!\bfp'_2\!=\!\half\bfpp_2\, . \label{dd4}\ee We also
factor the value ${\rm Tr}\{\brs\,\brq\,\bre\}$ from $\omega$ and
$\rho$-exchange amplitudes out of integrals~(\ref{dd3}), fixing
the momentum $\bfs$ according to prescription~(\ref{dd4}). Then,
$\bfs\!=\!\bfk\!+\half(\bfpp_1\!-\!\bfpp_2)\!=\!\half(\bfk\!+\!\bfq)$,
and we replace ${\rm Tr}\{\brs\,\brq\,\bre\}\!\to\!\half{\rm
Tr}\{\brk\,\brq\,\bre\}$ in Eqs.~(\ref{dd3}).

\vspace{1mm}

The meson propagator $G_h(s)\!=\!(s^2\!-\!m^2_h\!+\!i0)^{-1}$ in
Eq.~(\ref{dd3}) can be written as \be
G_h(s)=-\left[\bfs^2\!+\!\frac{Q_0}{m}(\bfp'^{\,2}_1\!+\!\bfp^2_2)\!+
\!a^2_h
-\!i0\right]^{-1}\!,{~~}Q_0=q_0+\!T_{d1}\!-\!\varepsilon_d\, ,
{~~} a^2_h=\!m^2_h\!-\!Q^2_0\, . \label{dd5}\ee Here, $T_{d1}$ is
the kinetic CM energy of the initial deuteron,
$\bfs\!=-\!\bfp'_1\!-\!\bfp_2\,$, and $\,Q_0$ is the excess energy
in the process $\gamma d\!\to\! pnh$.  The term $(Q_0/m)
(\bfp'^{\,2}_1\!+\!\bfp^2_2)$ in Eq.~(\ref{dd5}) takes into
account the kinetic energies of the intermediate nucleons.  In
this Section, we neglect this term, \textit{i.e.}, we use a
``static" approximation.  However, in order to study the energy
dependence of the differential cross section of $\eta$ production
($Q_0\sim m_{\eta}$) in the threshold region, we shall restore
this term in the amplitude with intermediate $\eta$ meson in
Subsection~\ref{sec:nucl}.

In the ``static" approximation,
$G_h\!=\!-(\bfs^2\!+\!a^2_h-\!i0)^{-1}$.  $\bfr$-representation is
convenient to calculate the double-scattering amplitudes. Let us
introduce the Fourier transformations for $\bfs$-dependent parts
of the integrands~(\ref{dd3}): \be \frac{1}{\bfs^2\!+\!a^2_h
-\!i0}=\intesr h_{11}(r)\, , {~~~}
\frac{\bfs\,F(s)}{\bfs^2\!+\!a^2_h-\!i0}=\intesr \bfr\,
h_{12}(r)\, , \label{dd6}\ee
$$
\frac{s_i s_j}{\bfs^2\!+\!a^2_h-\!i0}\,F^2(s)
=\intesr [r_i r_{j\,} h_1(r) +\dij r^2 h_2(r)]\, .
$$
The functions $h_{11,12,1,2}(r)$ are given in Appendix~\ref{sec:func}.
Let us also define the integrals
\be
\int f^2(r)\, h_{11}(r)=A_{11}\, , {~~~~}
\int f^2(r)\, h_{12}(r)\,\bfr =\bfm\,A_{12}\, ,
\label{dd7}\ee
$$
\int f^2(r)\,h_1(r)\,r_i r_j =m_i m_j\,A_{221} +\dij\,A_{222}\, ,
{~~~} \int f^2(r)\,r^2\, h_2(r)=A_{223}\, ,
$$
where $f(r)\!=\!u(r)/(r\sqrt{2})$, and $u(r)$ is the $s$-wave part
of DWF.   For the integrals in Eqs.~(\ref{dd7}), we use a
short-hand notation \be \int\,...\equiv\inteqr\,...{~}, {~~~}
\bfqq=\half (\bfk+\bfq)\, , {~~~} \bfm=\frac{\bfqq}{Q}\, .
\label{dd8}\ee The expressions for $A_{11\,}$, $A_{12\,}$,
$A_{221\,}$, $A_{222\,}$, and $A_{223\,}$ are given in
Appendix~\ref{sec:int2}. Rewriting integrals~(\ref{dd3}) in
$\bfr$-representation using Eqs.~(\ref{ss5}), (\ref{dd6}), and
(\ref{dd7}), we obtain the following expressions for amplitudes
$M_{\,ihj}$:

\vspace{-4mm}
\begin{eqnarray}
\rptt M_{1h1} \!\!&=&\! y_{1h1\,} A_{11}\,
{\rm Tr}\{\brepb\bre\,\brepa\} {~~~~~} (h=\pi, \eta)\, ,
\label{dd9} \\
\nonumber
\rptt M_{1\pi 2} \!\!&=&\!y_{1\pi 2\,} A_{12}\,
{\rm Tr}\{\brepb\bre\,\brepa\brm\brk\}\, , {~~~~~~~} M_{2\pi 1}\!
=\!y_{2\pi 1\,} A_{12}\,{\rm Tr}\{\brepb\brm\brq\,\bre\,\brepa\}\, , \\
\nonumber
\rptt M_{2\pi 2}\!\!&=&\!y_{2\pi 2\,} \left[A_{221}\,{\rm Tr}
\{\brepb\,\brm\,\brq\,\bre\,\brepa\brm\,\brk\} + (A_{222}+A_{223})\,
{\rm Tr}\{\brepb\,\bfsig\,\brq\,\bre\,\brepa\bfsig\,\brk\}\,\right]\, ,
\\
\nonumber
\rptt M_{\omega\pi 1} \!\!&=&\!y_{\omega\pi 1\,} A_{11}\,
(\bfeps_{1\,}\bfeps^*_2)\,{\rm Tr}\{\brk\brq\,\bre\}\, , {~~~~~}
M_{\omega\pi 2}\!=\!\half\,y_{\omega\pi 2\,} A_{12}\,
{\rm Tr}\{\brepb\,\brepa\brm\brk\}\,{\rm Tr}\{\brk\,\brq\,\bre\}\, , \\
\nonumber
\rptt M_{\,\rho\eta 1} \!\!&=&\!y_{\,\rho\eta 1\,} A_{11}\,
(\bfeps_{1\,}\bfeps^*_2)\,{\rm Tr}\{\brk\brq\,\bre\}\, .
\end{eqnarray}

By summing the amplitudes from Eqs.~(\ref{ss4}) and (\ref{dd9}),
we obtain the total amplitude $\mgd$~(\ref{di1}).  Note that the
gauge invariance of this amplitude comes from the
Lagrangians~(\ref{a2}) and (\ref{b2}) and that it is not violated
by the nucleon off-shell effects in the deuteron.  For simplicity,
we do not take into account here VME terms in the double
scattering amplitudes, \textit{i.e.}, the amplitudes $M_{\omega\pi
1\,}$, $M_{\omega\pi 2\,}$ and $M_{\,\rho\eta 1}$ will be
neglected in the numerical calculations of
Section~\ref{sec:result}.  The square $\overline {|\mgd|^2}$, of
the total amplitude, averaged (summed) over initial (final)
polarizations, is rather cumbersome, and we do not write it here.

\section{Numerical results}
\label{sec:result}

\subsection{Nucleon kinetic energy terms}
\label{sec:nucl}

Before we consider the differential cross section of \reac\, let
us discuss the threshold effect of intermediate $\eta$ production
in the double scattering amplitude. The contributions of
intermediate $\eta$ meson to the reaction amplitude comes from the
terms $M_{1\eta 1\,}$ and $M_{\,\rho\eta 1}$~(\ref{dd9}) which
contain the integral $A_{11}$.  We shall recalculate $A_{11}$ in a
``non-static" case using the $\eta$ propagator~(\ref{dd5}) with
nucleon kinetic energy (NKE) terms
$\,(Q_0/m)(\bfp'^{\,2}_1\!+\!\bfp^2_2)\,$ taken into account, and
compare it with the result of the ``static" approximation, used in
Eqs.~(\ref{dd6}) and (\ref{dd7}).

To simplify the calculation of $A_{11}$ for the ``non-static"
case, let us replace the Bonn DWF by the effective Gaussian
$s$-wave function $\psi(r)=B\exp(-br^2)$. In order to fix the
slope parameter $b$, let us note that for \reac\ at the $\eta$
threshold with backward outgoing $\pi^0$, we have
$a^2_h\!=\!0\,(h\!=\!\eta)$ in Eq.~(\ref{dd5}) and $\bfqq\approx
0$ in Eqs.~(\ref{dd7}). On the other hand, we have ${\dist
A_{11}(a^2_h\!=\!Q\!=\!0)\sim <\!1/r\!>_{\!d}}$ from
Eqs.~(\ref{dd7}). We fix the value $b$ from the mean value
$<\!1/r\!>_d$ corresponding to the $s$-wave part of the Bonn DWF.
The expression of $A_{11}$ with the Gaussian DWF is given in
Appendix~\ref{sec:int3}.

Fig.~\ref{fig:g4} shows Re$_{\,}A_{11}$ and Im$_{\,}A_{11\,}$
calculated in the ``static" approximation, as a function of the
photon laboratory energy $\ega$ for several values of
$z=\cos\theta$, where $\theta$ is the CM scattering angle of the
outgoing $\pi^0$ meson.  Here, one can see that the results
obtained with Bonn (solid curves) and Gaussian (dashed curves) $s$
DWF are quite close to each other.  The function Re$_{\,}A_{11}$
peaks at $\ega\!=\!E_{th}\!\approx\!630$~MeV, where $E_{th}$ is
the threshold for the reaction $\gamma d\!\to\!\eta pn$, and
Im$_{\,}A_{11}\!=\!0$ at $\ega\!< \!E_{th\,}$.  Note that the left
(right) derivative
 $d_{\,}$(Re$_{\,}A_{11})/d\ega\to+\infty$
[$d_{\,}$(Im$_{\,}A_{11})/d\ega\to+\infty$] at $\ega\to E_{th}$.
These properties come from the ``static" approximation used in
Eqs.~(\ref{dd6}).  In the $\eta$ threshold region, the function
Im$_{\,}A_{11}$ should depend on $\ega$ as a 3-particle $NN\eta$
phase space, \textit{i.e.}, $\sim Q^{3/2}_0$, where $Q_0$ is the
excess energy given in Eqs.~(\ref{dd5}) and
$Q_0\!\approx\!\ega\!-\!E_{th}$. In fact, when NKE terms
$\,(Q_0/m)(\bfp'^{\,2}_1\!+\!\bfp^2_2)\,$ in the $\eta$
propagator~(\ref{dd5}) are neglected, then Im$_{\,}A_{11}$ behaves
as a 2-particle phase space, \textit{i.e.}, $\sim Q^{1/2}_0\,$.

\begin{figure}[th]
\centering{
\includegraphics[height=0.5\textwidth, angle=90]{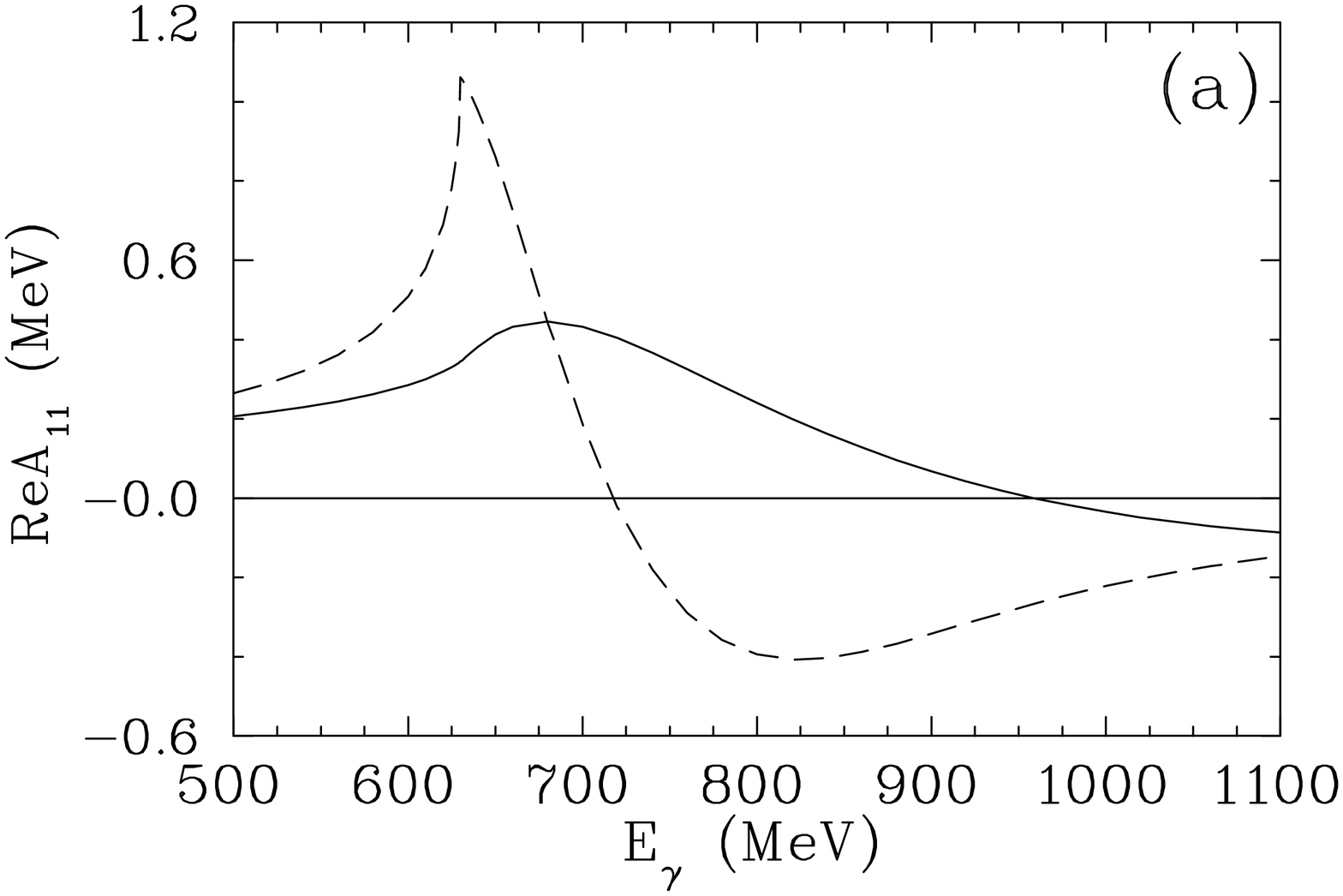}\hfill
\includegraphics[height=0.5\textwidth, angle=90]{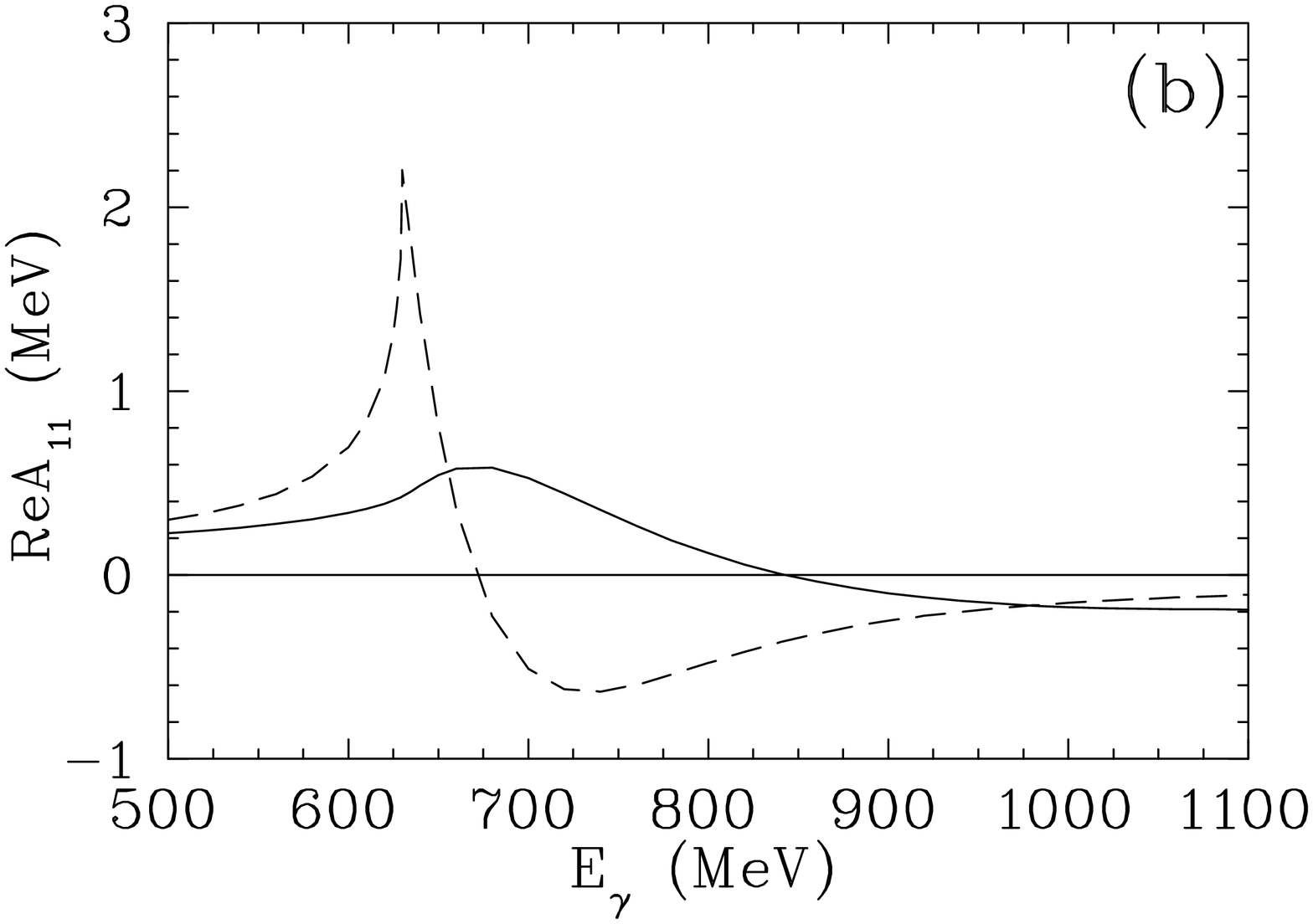}
\includegraphics[height=0.5\textwidth, angle=90]{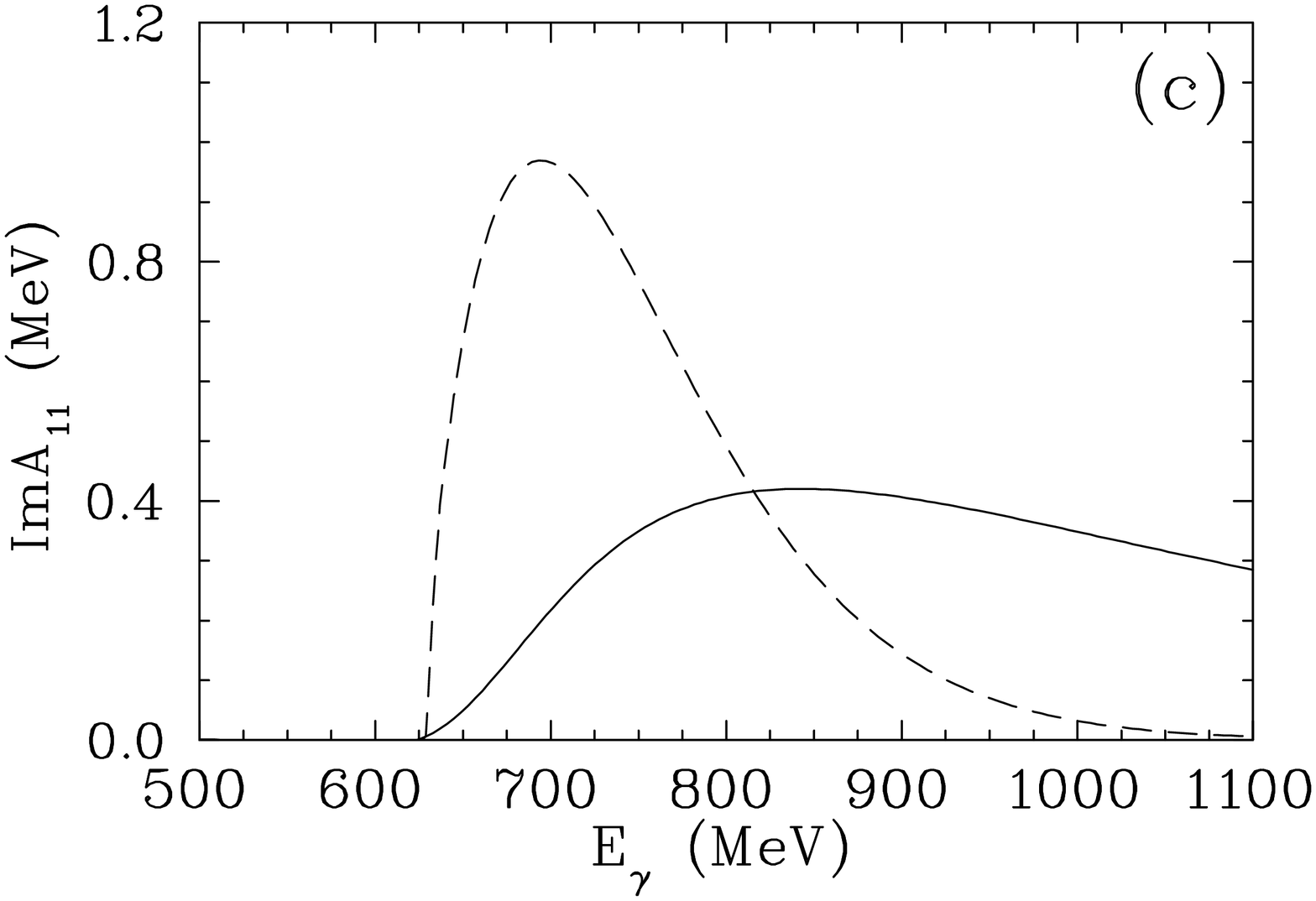}\hfill
\includegraphics[height=0.5\textwidth, angle=90]{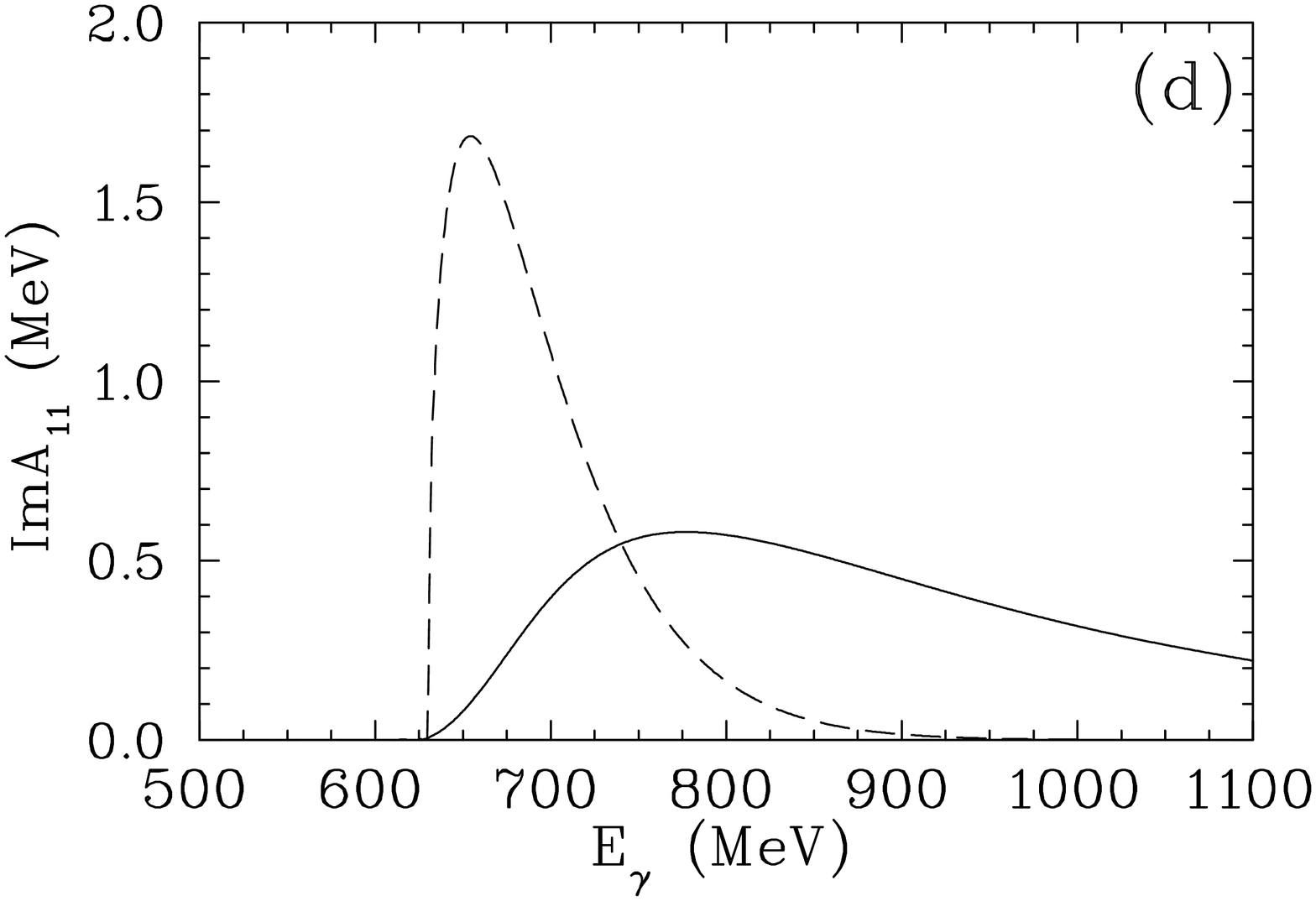}
\includegraphics[height=0.5\textwidth, angle=90]{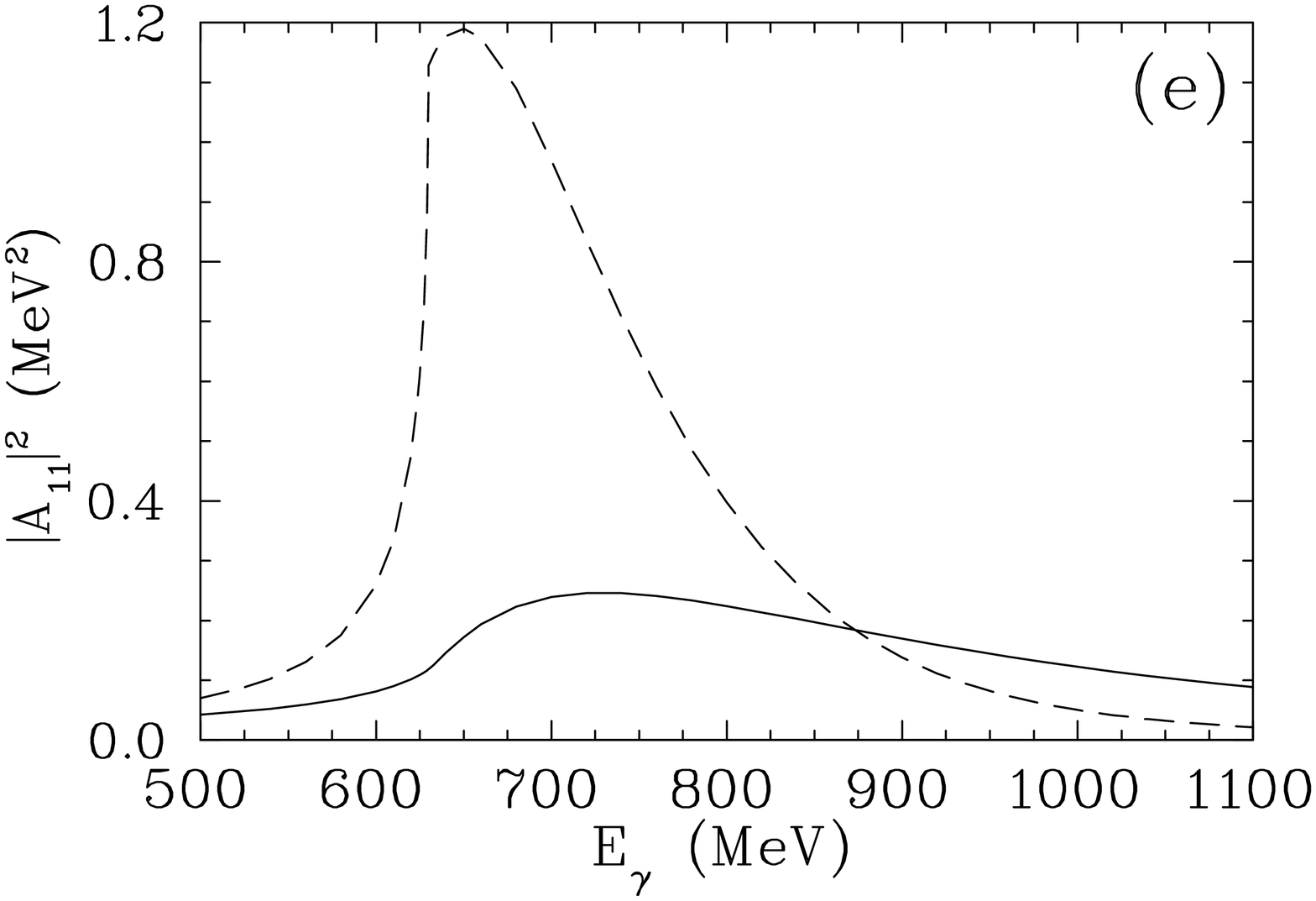}\hfill
\includegraphics[height=0.5\textwidth, angle=90]{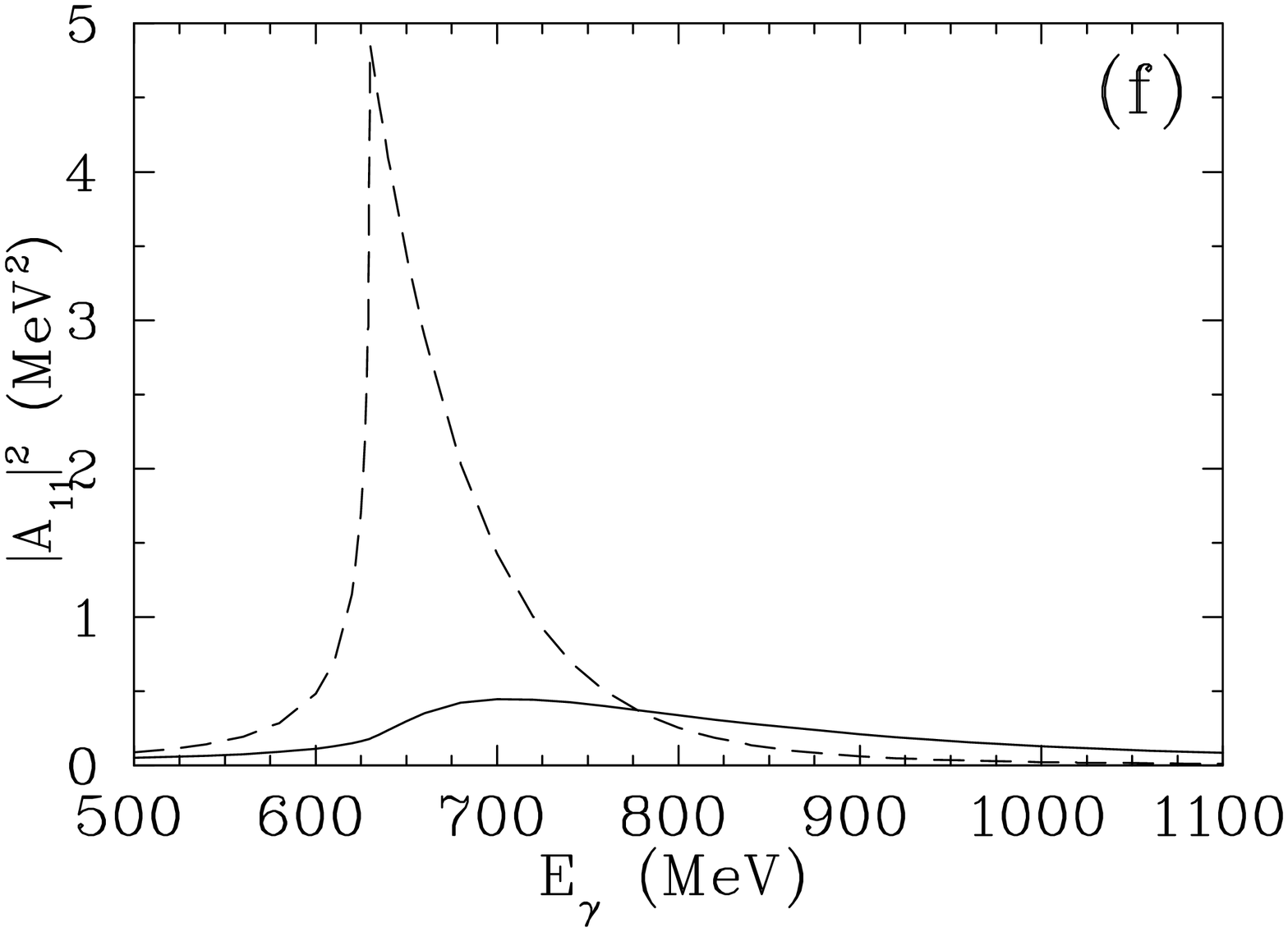}
}\caption{The results for (a) and (b) Re$\,A_{11}$;
          (c) and (d) Im$\,A_{11}$; and (e) and (f)
          $|A_{11\,}|^2$ at (a), (c), and (e)
          $z\!=\!-0.55$; and (b), (d), and (f)
          $z\!=\!-0.85$ with the Gaussian $s$ DWF.
          Solid (dashed) curves correspond to
          ``non-static" (``static") case.
          \label{fig:g5}}
\end{figure}

In Fig.~\ref{fig:g5}, we compare Re$_{\,}A_{11\,}$,
Im$_{\,}A_{11\,}$, and $|A_{11}|^2$ calculated with the Gaussian
DWF in ``static" (dashed curves) and ``non-static" (solid curves)
cases. Results are shown for two values $z=-0.55$
(Fig.~\ref{fig:g5}a,c,e) and $z=-0.85$ (Fig.~ \ref{fig:g5}b,d,f).
In the ``non-static" case, Im$_{\,}A_{11\,}\sim Q^{3/2}$ in the
small region close to the $\eta$ threshold and then due to DWF
begins to decrease where $\ega$ increases.  The energy dependence
of $|A_{11}|^2$ clearly demonstrates the $\eta$-threshold effect
from the loop diagram (Fig.~\ref{fig:g1}b) in the energy behavior
of the differential cross section of \reac\ when all kinematical
factors from subreactions on the nucleons are neglected.
Fig.~\ref{fig:g5}e,f show that when NKE terms are included then
$|A_{11}|^2$ turns to be much smoother function instead of
exhibiting sharp peaking in the ``static" approximation case.

Finally, for the differential cross sections in
Subsection~\ref{sec:nucl}, all double-scattering amplitudes with
intermediate pion are calculated in a ``static" approximation
using the Bonn DWF~\cite{Bonn} and the expressions for the
integrals $A_{11,12,221,222,223}$ are given in
Appendix~\ref{sec:int2}.  For the amplitudes with an intermediate
$\eta$ meson, NKE terms are taken into account and we use $A_{11}$
given in Appendix~\ref{sec:int3}.

\subsection{Differential cross section of the reaction
$\gamma d\to\pi^0d$}
\label{sec:dsg}

The amplitude of \reac\ as expressed by Eqs.~(\ref{ss9}) and
(\ref{dd9}) depends on a number of parameters.  In the
Table~\ref{tbl1}, we list sets of helicity amplitudes
$A^{p,n}_{1/2}$ for photon couplings to spin-$\half$ resonances
(see Eqs. ~(\ref{a10})) used in our amplitudes.
\begin{table}[th]
\caption{$N(1535)$ and $N(1440)$ resonance couplings.
         Units are (GeV)$^{-1/2}\times 10^{-3}$.
         \label{tbl1}}
\begin{tabular}{|c c|c|c|c|c|c|c|}
\colrule
\rule{0pt}{16pt} & Ref. & \protect\cite{Arndt}-1 &
\protect\cite{Arndt}-2 &
\cite{CR} & \cite{Benme} & \protect\cite{Drech} &
PDG~\protect\cite{PDG} \\
\colrule
\rule{0pt}{16pt} \NN\
 & $A^p_{1/2}~$ & ~78 & ~50 & ~53 & 97 & ~67 &  ~90 \\
 & $A^n_{1/2}~$ & -50 & -37 & -98 & -- & -55 &  -46 \\
\colrule
\rule{0pt}{16pt} \PP\
 & $A^p_{1/2}~$ & -66 & -64 & -69 & -- & -71 &  -65 \\
 & $A^n_{1/2}~$ & ~50 & ~45 & ~56 & -- & ~60 &  ~40 \\
\colrule
\end{tabular}
\end{table}
We use the values $A^{p,n}_{1/2}$ from column ~\cite{Arndt}-1
(1-st variant from Ref.~\cite{Arndt}) which is approximately the
mean values among those given in Table~\ref{tbl1}.

For the partial ($\Gamma_{ih}$) and total ($\Gamma_i$)
widths~(\ref{a8}) of \NN\ and \PP\ at nominal masses
($\sqrt{s}\!=\!m_{1,2}$), we use the values
\be
\Gamma_{1\pi}=\Gamma_{1\eta}=0.5\,\Gamma_1\, , {~~~}
\Gamma_{2\pi}=0.65\,\Gamma_2\, ,{~~~}
\Gamma_1\!=150~{\rm MeV}\, , {~~~}
\Gamma_2\!=350~{\rm MeV}\, ,
\label{cs1}\ee
and take $\Lambda\!=1$~GeV in the form factor $F$~(\ref{a8})
of the hadronic $N(1440)$ decay.

The coupling constants $G_{\omega\pi\gamma}\,$ and
$G_{\rho\eta\gamma}$ are obtained through Eq.~(\ref{b3}) from
radiative widths~\cite{PDG}
\be
\Gamma_{\omega\!\to\pi^0\gamma}=8.7\cdot 10^{-2}\,\Gamma_{\omega}\, ,
{~~}\Gamma_{\rho^0\!\to\eta\gamma}=7.9\cdot 10^{-4}\,\Gamma_{\rho}\,
{~~} (\Gamma_{\omega}=8.44~{\rm MeV},
{~~} \Gamma_{\rho}\approx 150~{\rm MeV}).
\label{cs2}\ee

The strong coupling constants $g_{\omega}\!=\!g_{\omega NN}$ and
$g_{\rho}\!=\!g_{\rho NN}$ are not well determined as it was
mentioned in Ref.~\cite{Drech}. In various analyses, they vary in
the ranges~\cite{David,Dumbra} $8<g_{\omega}<20$ and $1.8<
g_{\rho} <3.2$ (we discuss only the vector coupling constants,
while tensor couplings are neglected in our approximation
according to the results of Subsection ~\ref{sec:VME}).  Here we
list some values of these constants from the above papers: \be
g_{\omega}\!=\!21, {~}g_{\rho}\!=\!2~[5];{~~~~}
g_{\rho}\!=\!\frac{g_{\omega}}{3}\!=\!2.66~[4];
{~~~~}\frac{g^2_{\rho}}{4\pi}\!=\!0.84, {~}
\frac{g^2_{\omega}}{4\pi}\!=\!20~[18]\,({\rm full~model}).
\label{cs3}\ee
For the results, shown below, we take the value $g_{\omega}$ from
Ref.~\cite{Bonn}, \textit{i.e.}, $g^2_{\omega}/4\pi\!=\!20\,$
($g_{\omega}\approx 15.85$).  The constant $g_{\rho}$ is not used
in our calculations since $\rho$-exchange terms are neglected, as
was mentioned in Subsections~ \ref{sec:DandN} and~\ref{sec:VME}.

\begin{figure}[th]
\centering{
\includegraphics[height=0.5\textwidth, angle=90]{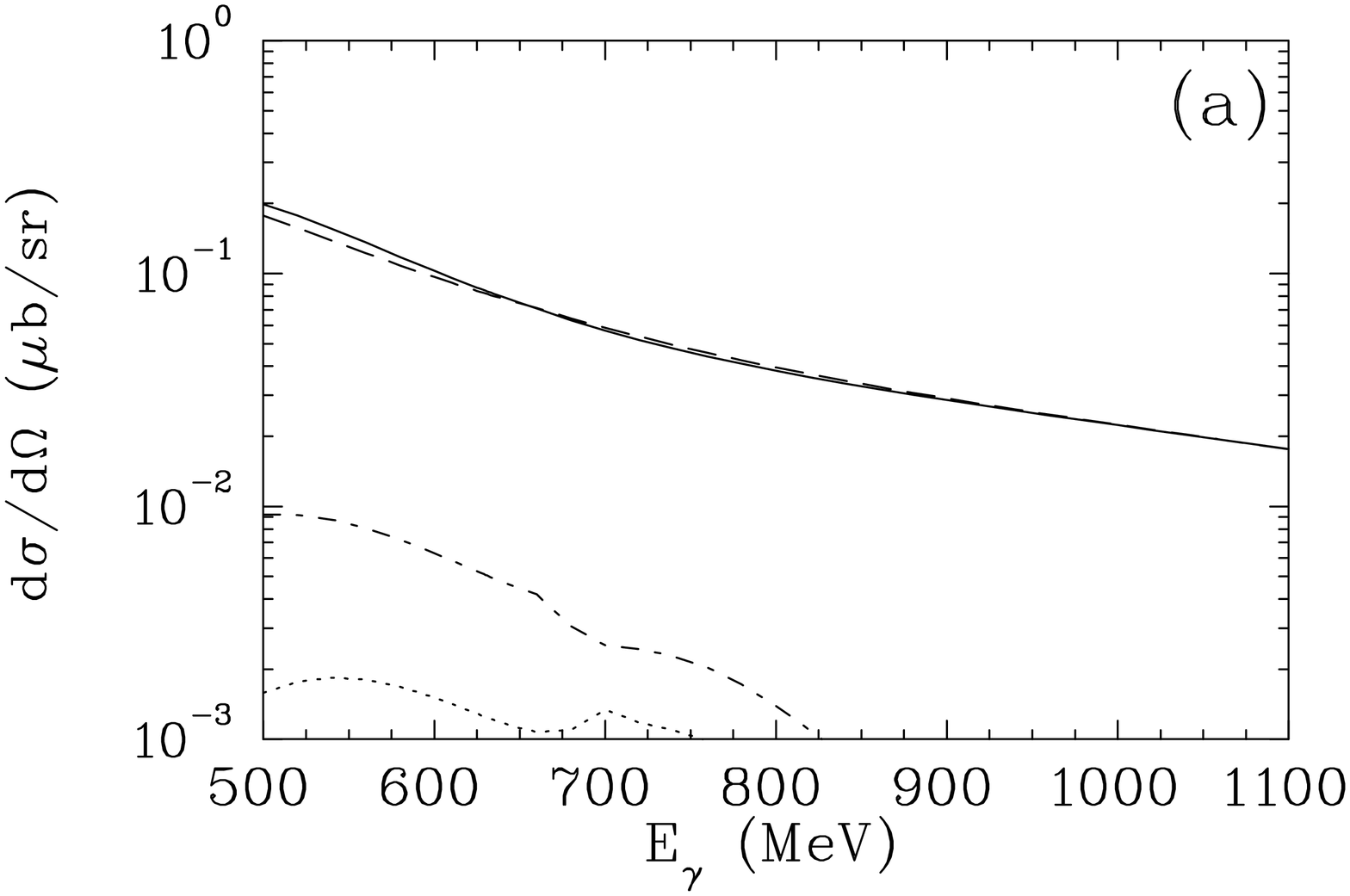}\hfill
\includegraphics[height=0.5\textwidth, angle=90]{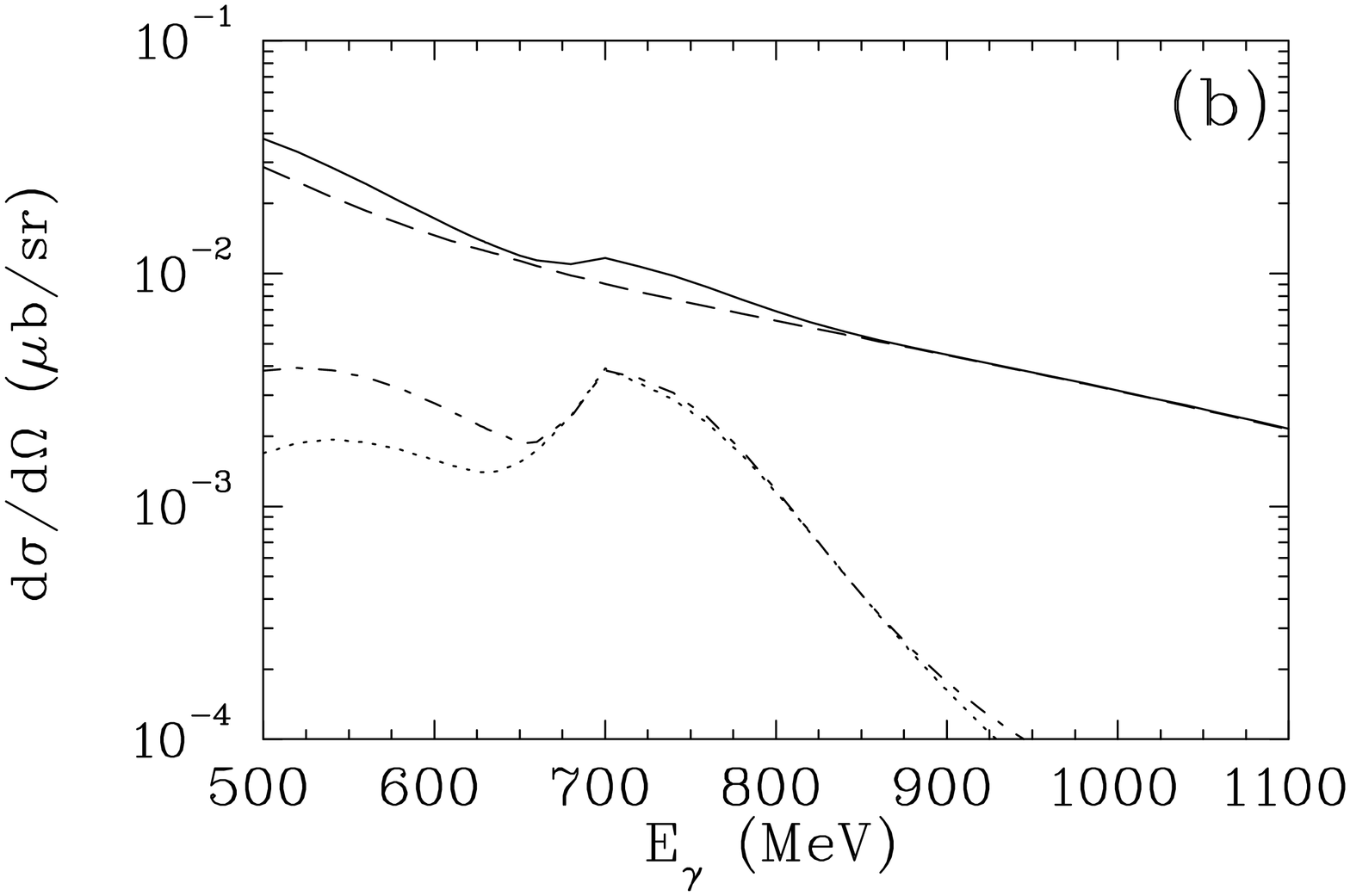}
\includegraphics[height=0.5\textwidth, angle=90]{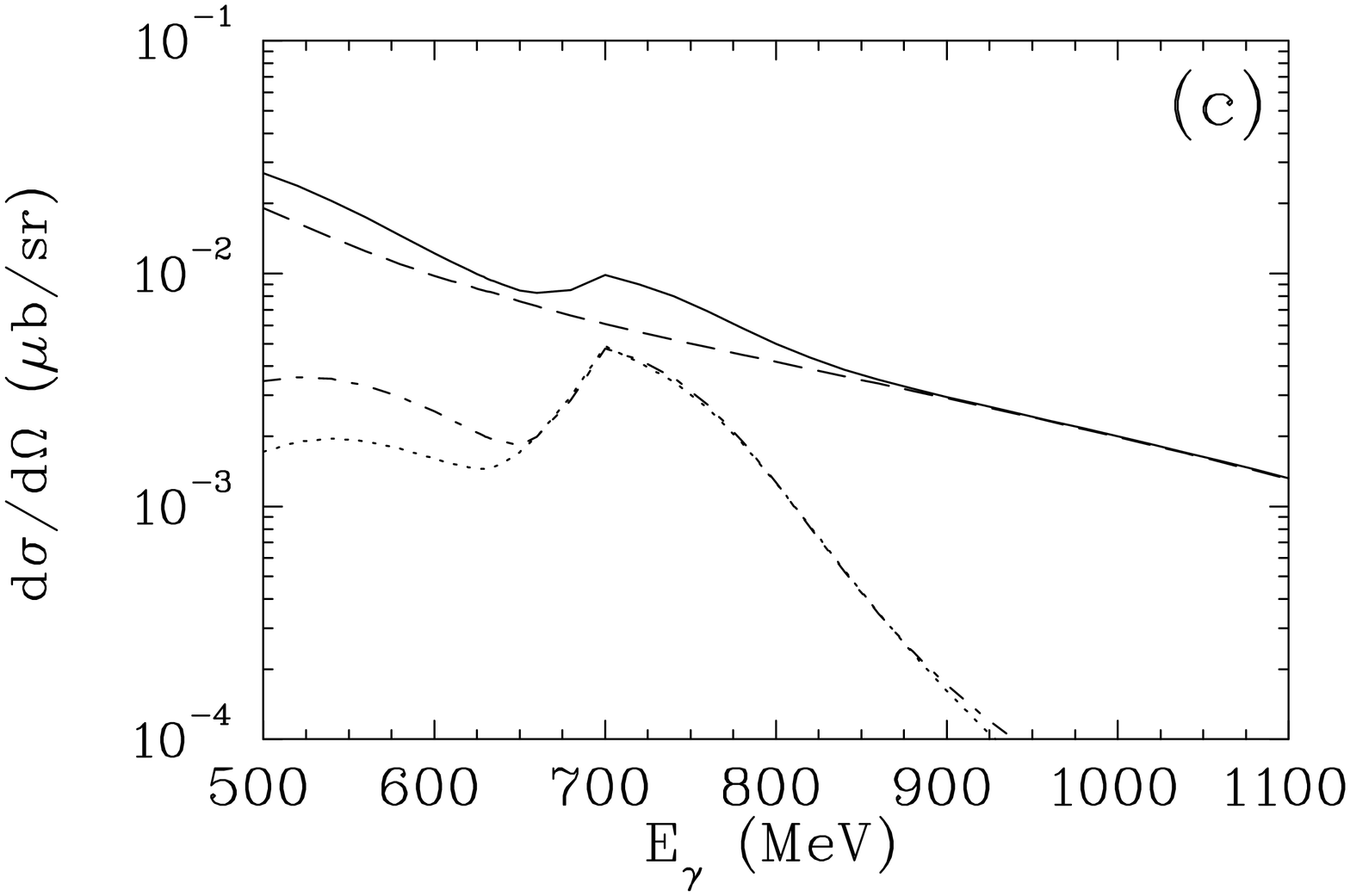}\hfill
\includegraphics[height=0.5\textwidth, angle=90]{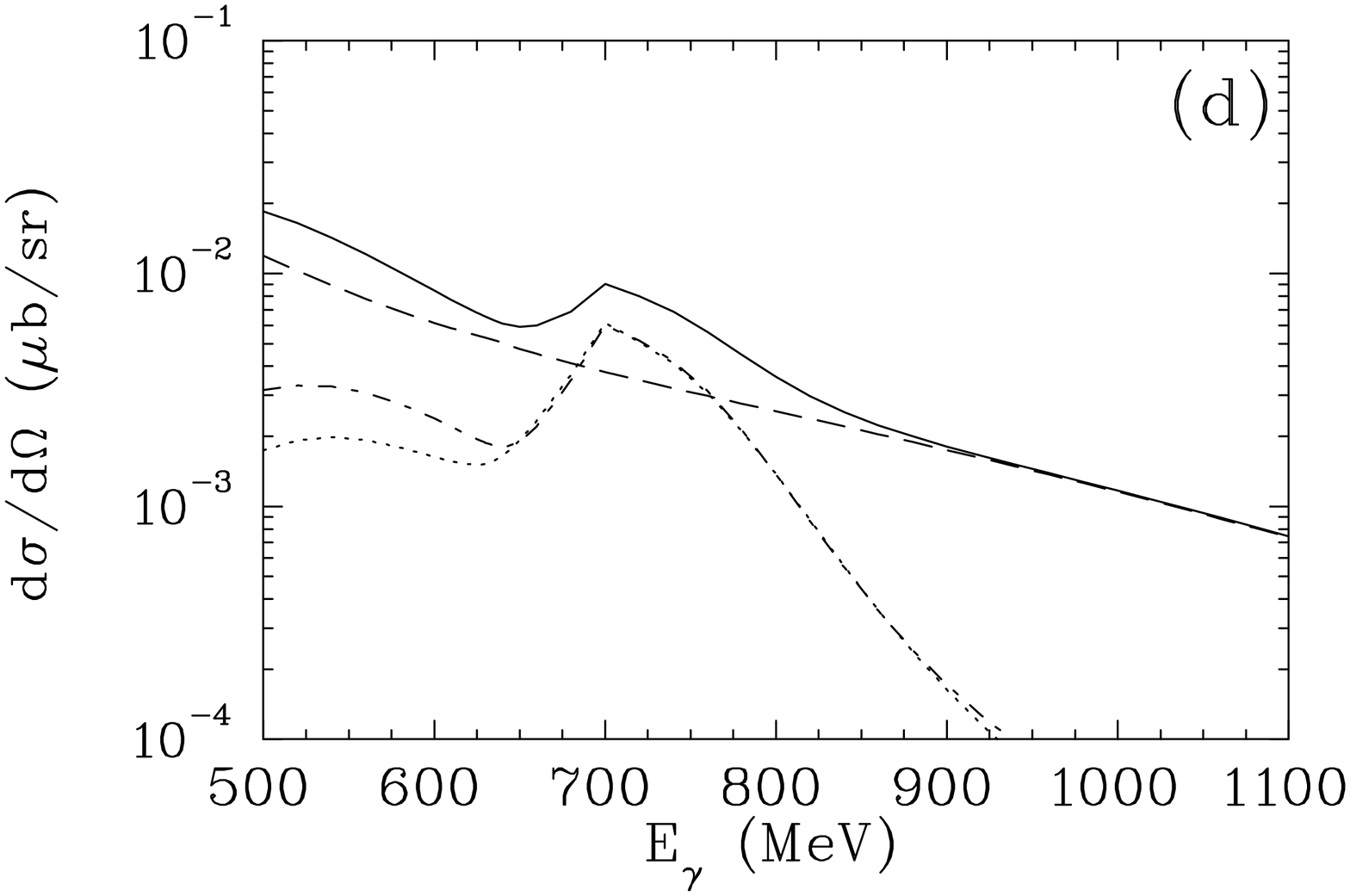}
\includegraphics[height=0.5\textwidth, angle=90]{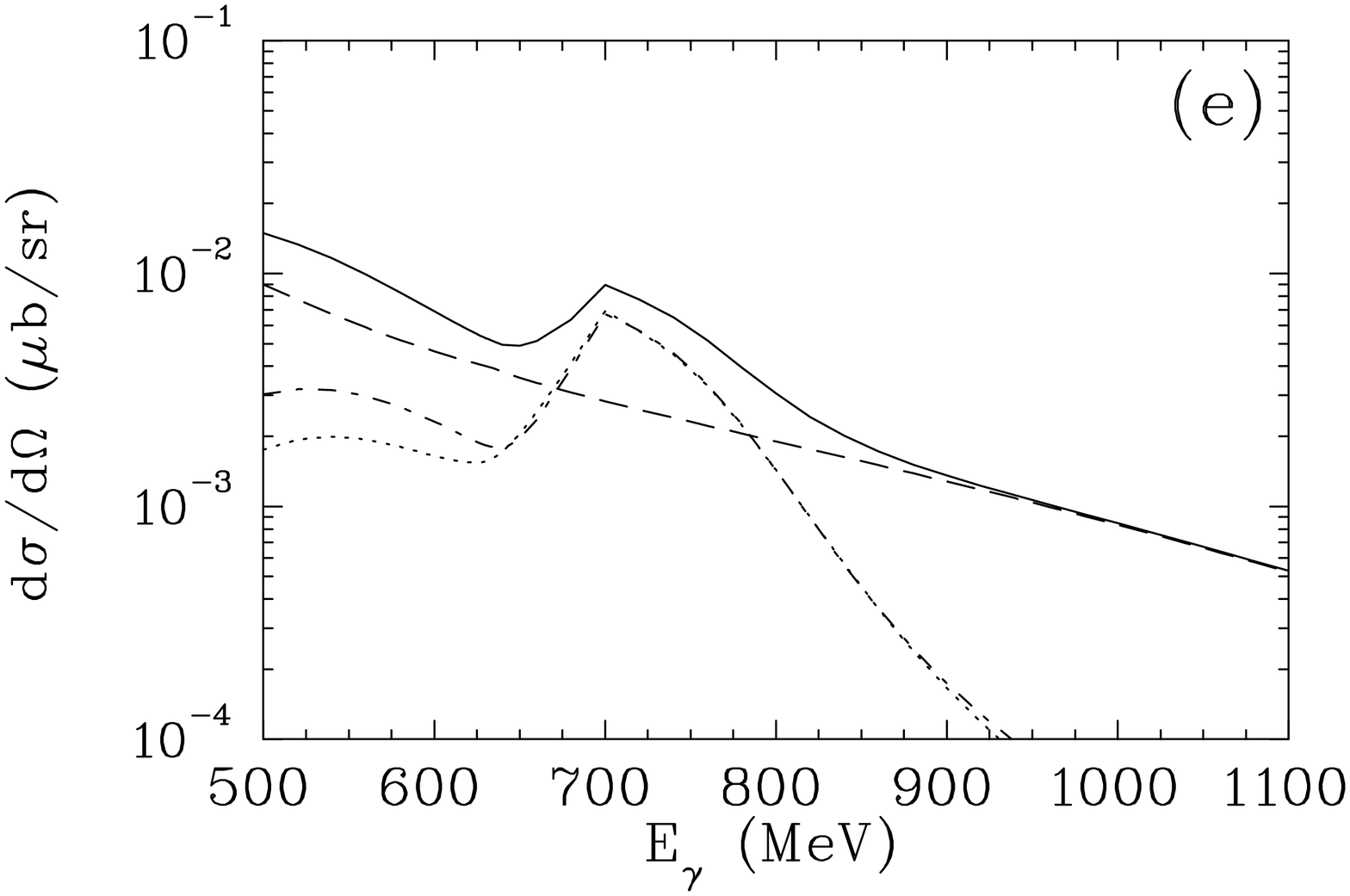}\hfill
\includegraphics[height=0.5\textwidth, angle=90]{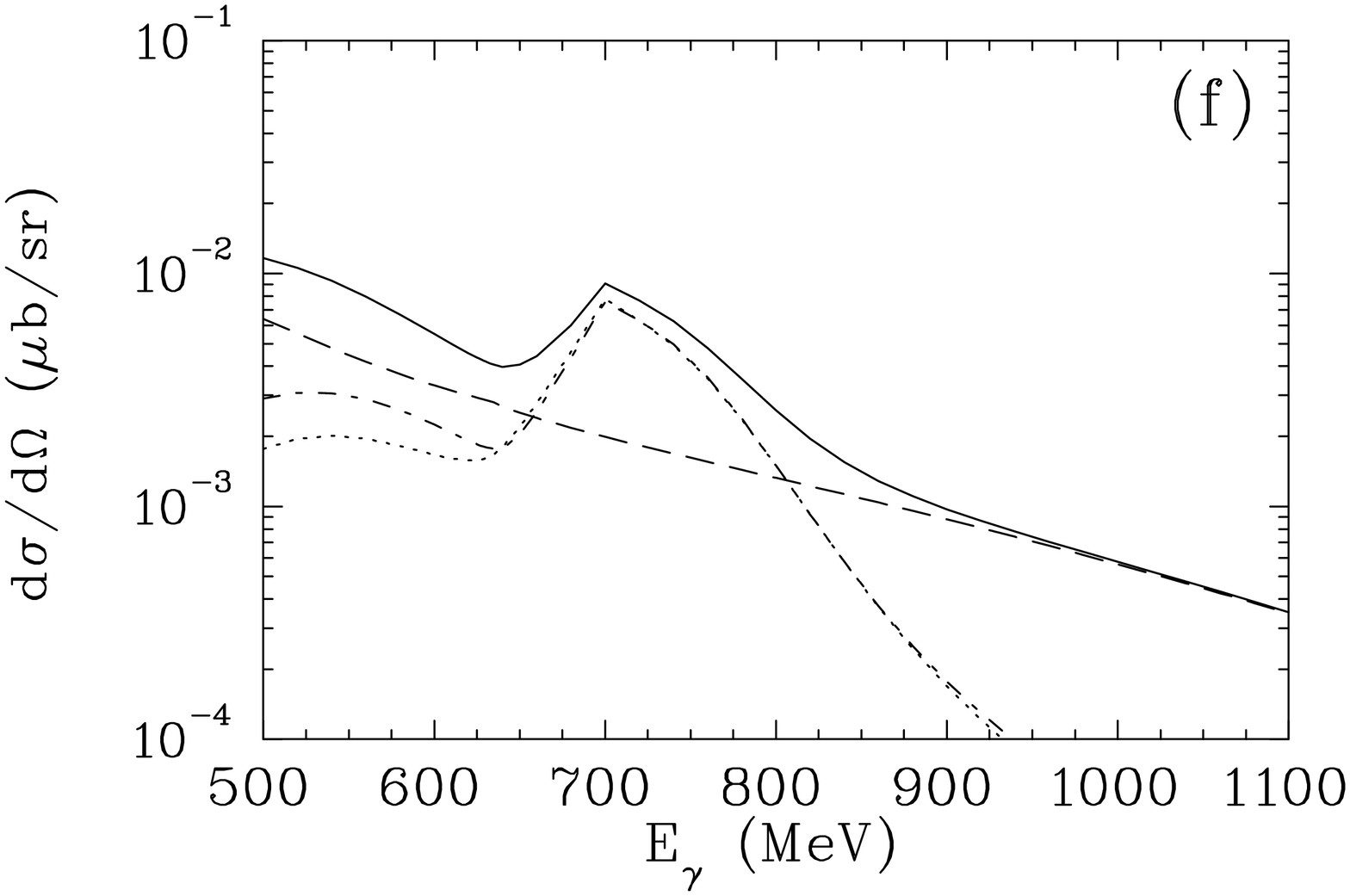}
}\caption{Differential cross sections of
          \reac\ vs photon laboratory energy $\ega$ at
          several values of $z\!=\!\cos\theta$.  Dashed
          (dotted) and solid curves show the contributions
          of single (double) - scattering amplitudes and total
          amplitude, respectively.  Dash-dotteded curves
          correspond to the contributions of total
          amplitude without $\omega$-exchange term.  The
          results are obtained with energy-dependent total
          width of \NN. (a) $z=0$, (b) $z=-0.55$, (c)
          $z=-0.65$, (d) $z=-0.75$, (e) $z=-0.8$, and
          (f) $z=-0.85$,
\label{fig:g6}}
\end{figure}

\vspace{2mm} In Fig.~~\ref{fig:g6}, we show the calculated
differential cross section of \reac\ with backward $\pi^0$
photoproduction as a function of the photon laboratory energy
$\ega$ at several fixed values of $z=\cos\theta$ from $z\!=\!0$ up
to $z\!=\!-0.85$ (the experimental CLAS data which are presented
in Ref.~\cite{CLAS} are at the same values $z$).  The results
shown by the solid curves obtained with the total amplitude
$\mgd\,$ consisting of the terms~(\ref{ss4}) and (\ref{dd9})
(without $M_{\omega\pi 1\,}$, $M_{\omega\pi 2}$, and $M_{\rho\eta
1}$ terms).  The other curves are explained in the figure caption.
One can see that the contribution from single-scattering
amplitudes dominates the cross section for $z\!=\!0$.  The
relative contribution from the other amplitudes increases as $z$
approaches $-0.85$. Note that the $\omega$-exchange amplitude
$M_{\omega}$ dominates in the total contribution from
single-scattering amplitudes.

In Fig.~\ref{fig:g6}, we see a maximum in the energy spectra of
the differential cross section at $\ega\approx 700$~MeV, in the
angular region of $z<-0.65$. This maximum is getting more
pronounced at $z\!\to\!-0.85$. The CLAS experimental
data~\cite{CLAS} also show the excess of events, but less sharp,
of the same order of magnitude (increasing at $z\!\to\!-0.85$) in
a region around the same energy for the same angles.  For more
detailed discussion, we should mention that the effective energies
$\,\sqrt{s}\,$ in subprocesses on the nucleons in the
double-scattering diagram (Fig.~\ref{fig:g6}b) are calculated in
the approximation~(\ref{dd4}). These energies $\,\sqrt{s}\,$ are
equal to the $\eta N$-threshold mass at $\ega\approx 700$~MeV. The
maxima in Fig.~~\ref{fig:g6} reflect the $\eta N$-threshold effect
from the \NN-propagators in the elementary amplitudes because we
use the energy-dependent width $\Gamma_1(s)$ of \NN\ according to
Eqs.~(\ref{a8}) and $\Gamma_{1\eta}(s)\!=\!0$ at $\,\sqrt{s}\le
m\!+\!m_{\eta}$ ($\ega\le 700$~MeV). The main effect comes from
the double-scattering amplitude $M_{1\eta 1}$ with two
\NN-propagators while the contribution from $M_{1\pi 1}$ is much
smaller due to a small $N(1440)N\pi$-coupling constant
($g_{1\pi}\ll g_{1\eta}$).  Thus, Fig.~\ref{fig:g6} demonstrates
the effects of the two-particle ($\eta N$) threshold in the
elementary amplitudes on the intermediate nucleons. At the same
time for the reasons discussed in Subsection~\ref{sec:nucl}, we do
not see a sizable threshold effect from the three-particle ($\eta
NN$) intermediate state ($E_{th}=630$~MeV).

Note that the prescription~(\ref{dd4}) usually works well due to a
rapid momentum dependence of the DWF in comparison with ones for
the amplitudes of the reactions on the nucleon. However, this
approximation does not reproduce adequately any sharp
peculiarities of elementary amplitudes in the amplitude of a
nuclear reaction.  Due to ``Fermi motion" within the deuteron, the
$\eta N$ threshold in the diagrams in (Fig.~~\ref{fig:g1}) are not
positioned at a fixed value of $\ega$. These effects tend to be
spread over some region of the incoming photon energy.  Thus, the
sharp maximum at $\ega\approx 700$~MeV in Fig.~\ref{fig:g6} should
be smoothed, but we have no proper simple procedure to do this.

\begin{figure}[th]
\centering{
\includegraphics[height=0.5\textwidth, angle=90]{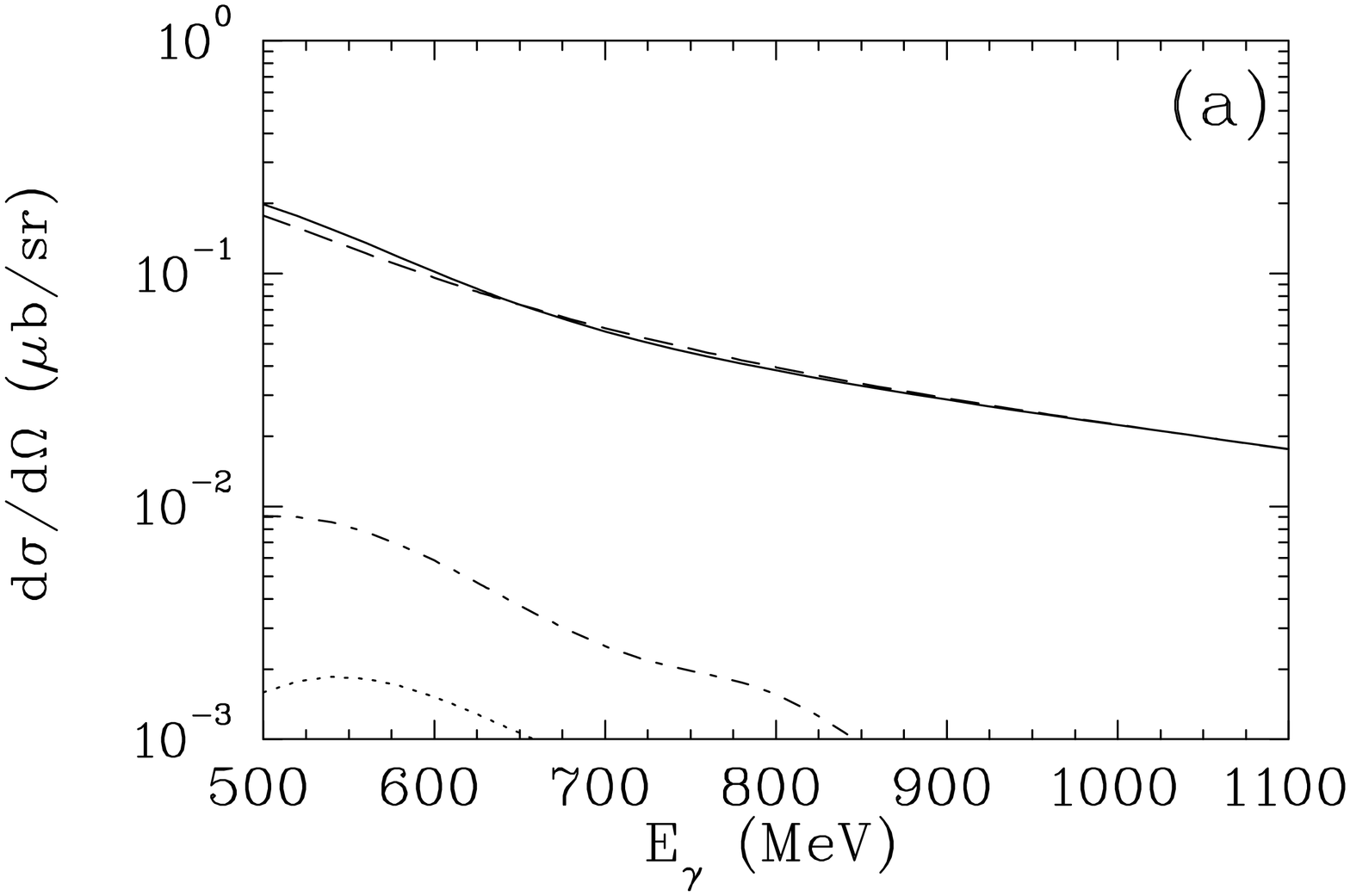}\hfill
\includegraphics[height=0.5\textwidth, angle=90]{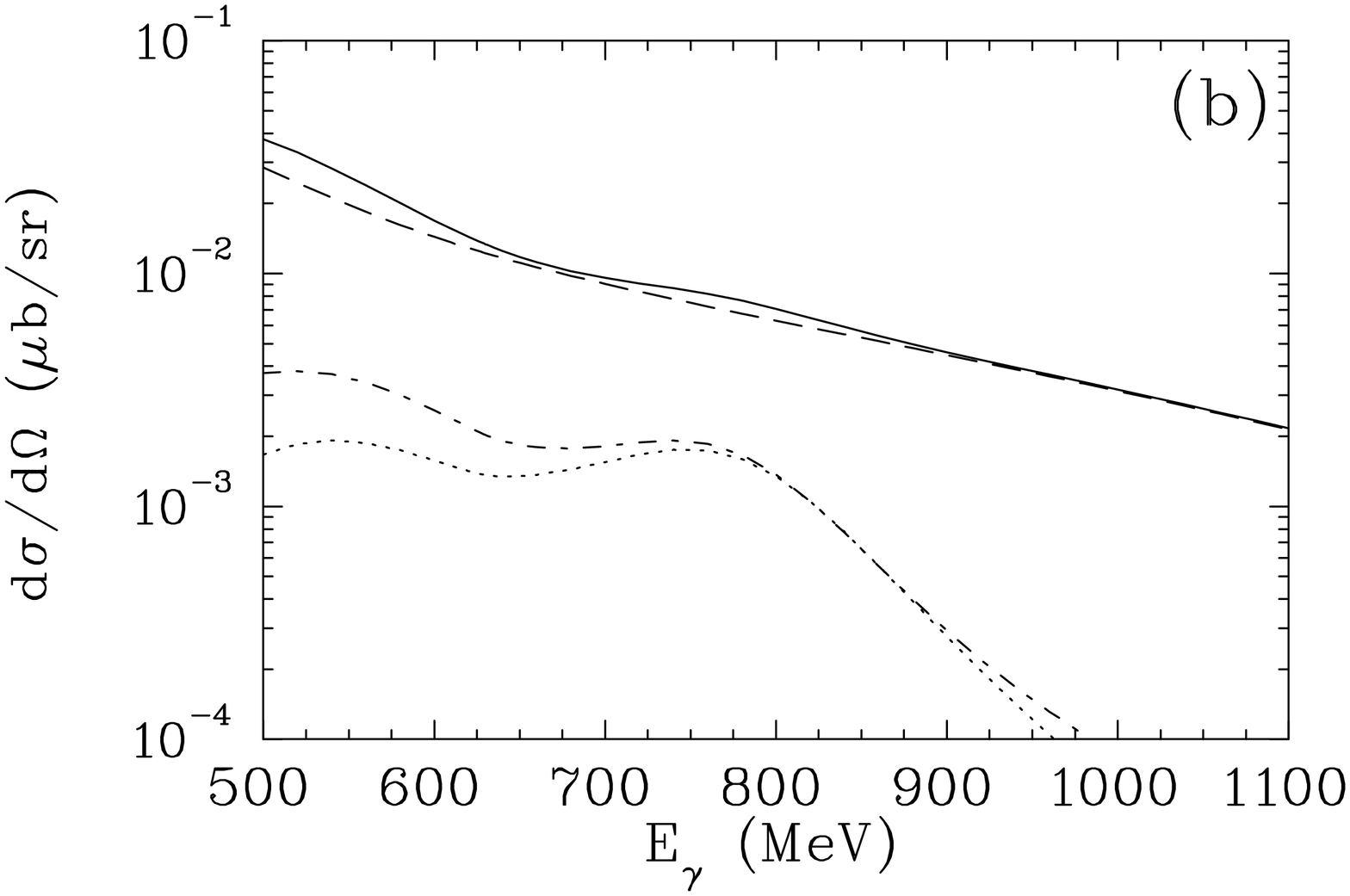}
\includegraphics[height=0.5\textwidth, angle=90]{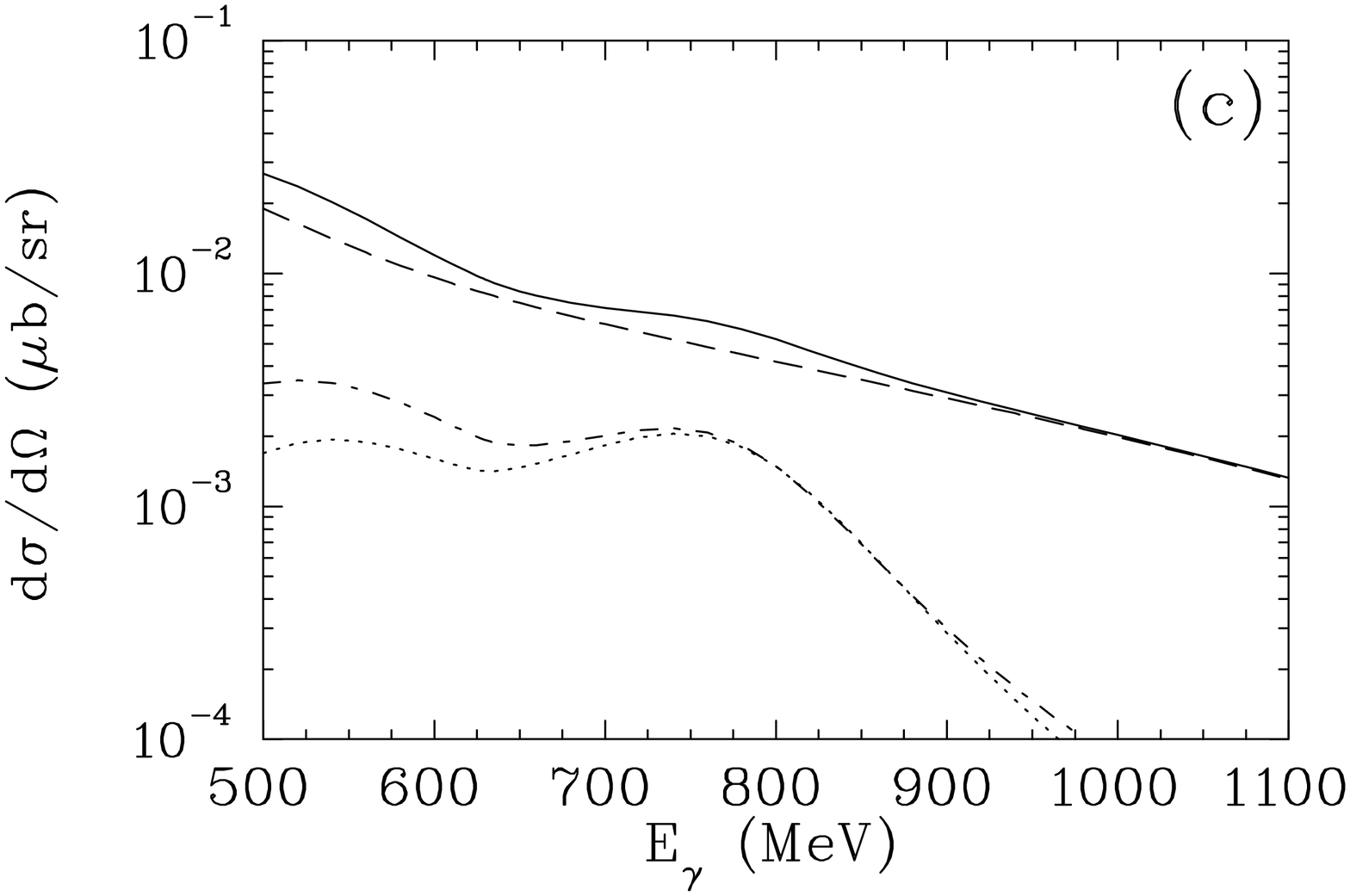}\hfill
\includegraphics[height=0.5\textwidth, angle=90]{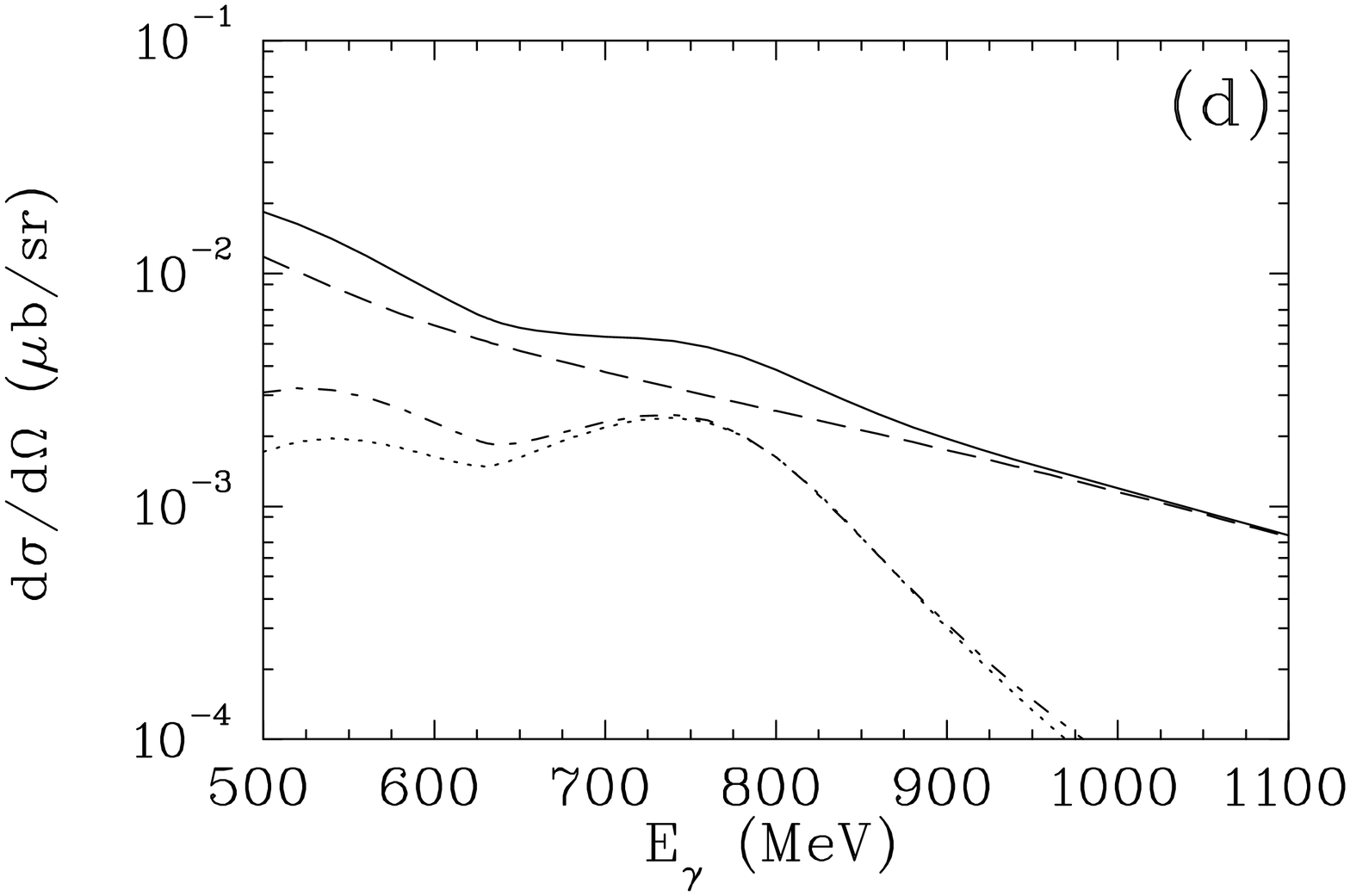}
\includegraphics[height=0.5\textwidth, angle=90]{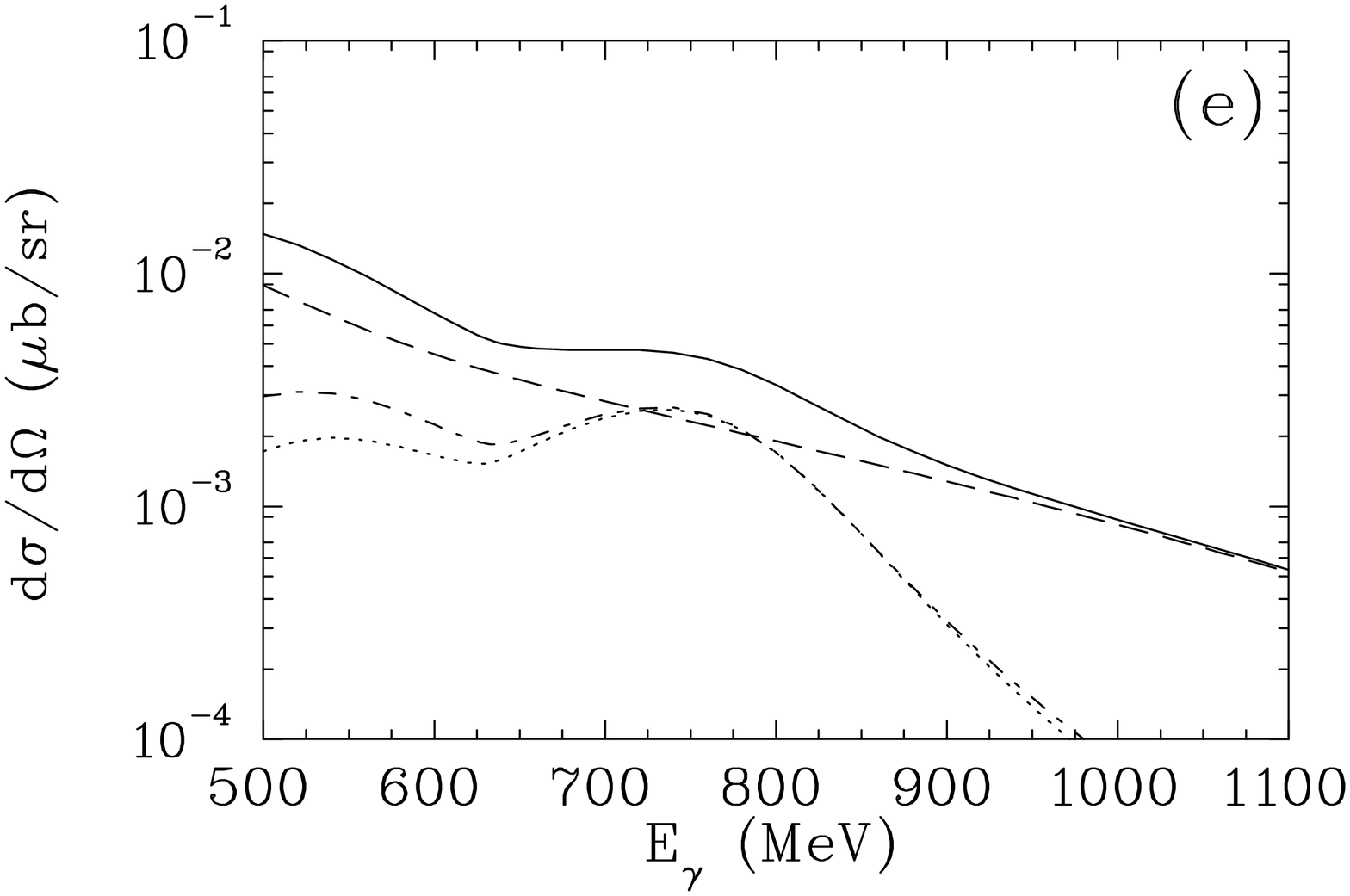}\hfill
\includegraphics[height=0.5\textwidth, angle=90]{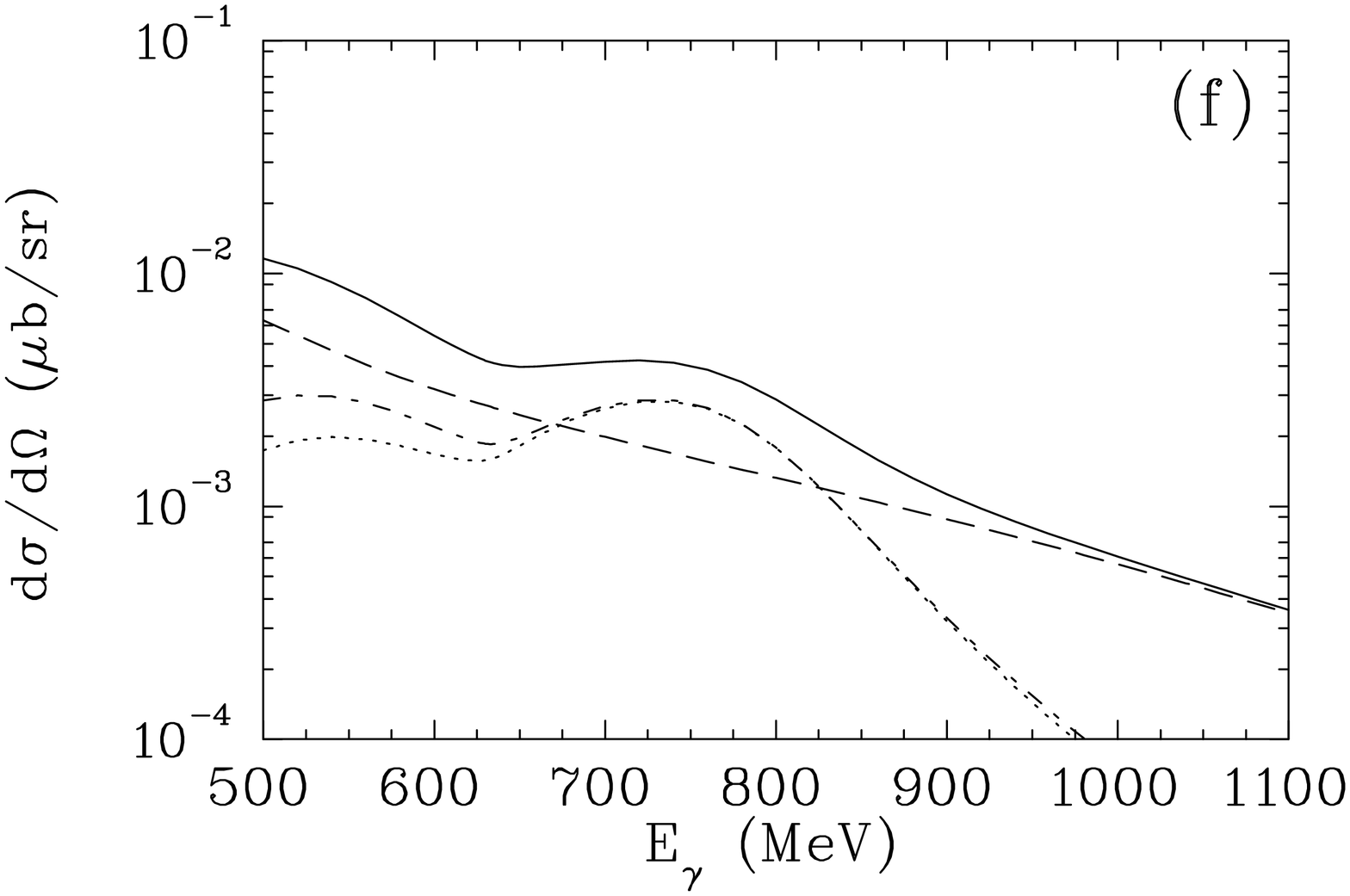}
}\caption{The same as in Fig.~\protect\ref{fig:g6} but
          the results are obtained with constant total
          width of \NN. \label{fig:g7}}
\end{figure}

Let us consider the case, when the \NN\ width in the elementary
amplitudes is a constant nominal value $\Gamma_1(s)\equiv
\Gamma_1=150$~MeV. In this case, shown in Fig.~~\ref{fig:g7}, we
have no $\eta N$-threshold effects in the elementary amplitudes
and no sharp maxima at $\ega\approx 700$~MeV.  Here, we see a
broad enhancements centered around $\ega=750$~MeV. These
enhancements appear mainly from the contribution of the amplitude
$M_{1\eta 1}$ with two \NN\ propagators. Their position is shifted
to the left of $\ega\approx 785$~MeV (laboratory photon energy on
the nucleon target with an effective CM energy equal to the \NN\
mass) due to decreasing factors from the DWF's in \reac. Note that
the energy-dependent \NN\ width $\Gamma_1(s)$ at the $\eta N$
threshold ($\sqrt{s}\!=\!m+m_{\eta}$), where
$\Gamma_{1\eta}(s)\!=\!0$, is about two times smaller than the
nominal value $\Gamma_1$ of the \NN\ resonance. Thus, taking \NN\
with constant nominal width, we get smaller values of the
corresponding elementary amplitudes in the region close to $\eta
N$ threshold. Therefore, the differential cross sections in
Fig.~\ref{fig:g6} at $\ega\sim 700$~MeV are essentially enhanced
in comparison with Fig.~\ref{fig:g7}. Finally, we conclude that
the enhancements in the energy region near the $\eta$ threshold,
shown in Figs.~\ref{fig:g6} and ~\ref{fig:g7}, are mainly due to
the \NN\ contributions to the double-scattering diagram with an
intermediate $\eta$ production.

The absolute values of the calculated differential cross section
at $z\!=\!0$ (Figs.~\ref{fig:g6} and~\ref{fig:g7}) are in an
approximate agreement with the CLAS data~\cite{CLAS}. For larger
angles $\theta$, our results are getting lower in absolute values
in comparison with the data.  At $z\!=\!-0.85$
(Figs.~\ref{fig:g1}f and~\ref{fig:g7}f), the calculated
differential cross sections are several times smaller than the
experimental ones.  We expect that the contribution from the
single-scattering amplitude $M_{\omega}$ in our treatment is too
large.  On the other hand, the VME terms in the double-scattering
amplitudes (not included here), being added, may essentially
improve our predictions at $z\to -0.85$, where contribution from
the term $M_{\omega}$ is getting smaller.  Thus, we hope that
decreasing the elementary $\omega$-exchange amplitude (taking
smaller coupling constant $g_{\omega N\!N}$ and introducing form
factors) and including the double-scattering amplitudes
$M_{\omega\pi 1}$ and $M_{\omega\pi 2}$ (the contribution from
$M_{\rho\eta 1}$ should be much smaller), we may have a good
description of the experimental absolute cross sections at
$z\!=\!0$ and essentially improve our predictions at $z\!\to
-0.85$.

\section{Conclusion}
\label{sec:conc}

We considered the energy dependence of the differential cross
sections of the reaction $\gamma d\to\pi^0d$ in a wide energy
range around the $\eta$ photoproduction threshold at several
backward CM angles of the outgoing pion.  Our calculations are
based on a nonrelativistic diagramatical technique and take into
account single- and double-scattering amplitudes.  We conclude
that the contribution of the double-scattering with intermediate
production of an $\eta$ meson can explain the broad cusp
experimentally observed by the CLAS Collaboration~\cite{CLAS} in
the energy dependence of the differential cross section near the
$\eta$ threshold. Indeed, our calculations show that a broad
enhancement (with the width of the order of 100~MeV) appears in
the energy behavior of the differential cross section at large
scattering angles $\theta$ in the $\eta$-threshold region.  This
enhancement becomes more pronounced as $\theta$ increases, and the
magnitude of the effect is in a qualitative agreement with the
CLAS data.

We discussed the role of the 3-particle $NN\eta$-threshold effect
taking into account ``non-static" corrections to the
$\eta$-propagator.  Indeed, we found that a sharp energy behavior
of the differential cross section at the energy of the $\eta$
threshold calculated in a ``static" approximation was essentially
smoothed by taking into account ``non-static" corrections to the
$\eta$ propagator. Our calculation results show that the
enhancements in the energy dependence of the differential cross
section are, to a great extent, due to the \NN\ contributions in
the elementary amplitudes of the double-scattering diagram with
intermediate $\eta$ production.

Our predictions depend on a number of parameters, some of which
are not well established (\textit{e.g.} the constants $g_{\rho
N\!N}$ and $g_{\omega N\!N}$, form-factor parameters,
\textit{etc.}). Not all of the possible diagrams are considered in
our analysis.  Our calculations do not include VME terms in the
double-scattering amplitudes or any other resonance amplitudes
besides the \NN\ and \PP\ contributions. Thus, there is a way for
further improvements of our predictions.

\acknowledgments

This work was supported in part by the Russian RFBR Grant
No.~02--02--16465, U.~S.~Department of Energy Grant
DE--FG02--99ER41110 and The George Washington University Center
for Nuclear Studies. The author (I.~S.) acknowledges partial
support from Jefferson Lab and the Southeastern Universities
Research Association under DOE contract DE--AC05--84ER40150.

\appendix

\section{The integrals $A_1$, $B_{1,2}$, and
$C_{1,2}$,~Eqs.(\ref{ss7})}
\label{sec:int1}

We use DWF from Bonn potential~\cite{Bonn} given in
parametrization of
$$
u(q)\!=\!\sum_i \frac{c_i}{m^2_i\!+\!q^2}\, , {~~}
u(r)\!=\!\sum_i c_i e^{-m_i r} , {~~}\sum_i c_i\!=\!0\, , {~~}
w(q)\!=\!\sum_i \frac{d_i}{m^2_i\!+\!q^2}\, ,
\eqno {\rm (A.1)}
$$ $$
w(r)\!=\!\sum_i d_i e^{-m_i r}\!
\left(1\!+\!\frac{3}{m_i r}\!+\!\frac{3}{m^2_i r^2}\right)\, , {~~~}
\sum_i d_i\!=\!\sum_i d_i m^2_i\!=\!\sum_i\frac{d_i}{m^2_i}\!=\!0\, ,
$$
where $u(q),~w(q)~(u(r), and ~w(r))$ are $s$- and $d$-wave parts of
the DWF in \bfq(\bfr)-space (see Eqs.~(\ref{ss2}) and
(\ref{ss5})). Using a DWF of the type~(A.1), one can calculate the
integrals in $\bfr$-space with the help (after integrating over
the angle) of the formula
$$
\int\limits^{\infty}_0\!dr\sum^n_i c_i\frac{\exp(a_i r)}{r^{n_i+1}}
=\sum^n_i \frac{c_i\,a^{n_i}_i}{n_i\,!}\,[S_{n_i}-\ln a_i\, ]
{~~} (n_i\ge 0)\, ,{~~~}S_n=\sum^n_{k=1}\frac{1}{k}\, , {~~}S_0 =0\, ,
\eqno {\rm (A.2)}
$$
which is valid if this integral converges, \textit{i.e.},
Re$\,a_i< 0$ and the integrand is finite at $r\!\to\! 0$.  These
conditions are satisfied our case.

Using Eqs.~(\ref{ss7}), (\ref{ss8}), (A.1), and (A.2), we obtain
the expression for $A_1$ in the form (in this Appendix for
integrals, we use a short-hand notation given by Eqs.~(A.3))
$$
A_1=\int f^2_1(r)=J_1\!+\!J_2\!+\!J_3 {~~~~~~~}
\left(\int\, ...\,=\intedr\, ...\right) , \eqno {\rm (A.3)}
$$

\vspace{-7mm}
\begin{eqnarray}
\nonumber
J_1\!\!&=&\!\!\int \frac{u^2(r)}{2r^2}\, = {~}
\scij{8\pi\Delta}\,\Aij\,
{~~~~~} (m_{ij}=m_i+m_j) \, , \\
\nonumber
J_2\!\!&=&\!\!\int \frac{u(r) w(r)}{\sqrt{2}\,r^2} =
\sum_{i\,j}\frac{c_{i\,} d_j}{8\sqrt{2}\,\pi}\,
\left[\,\frac{3m_i}{m^2_j}\,\Lij +
\frac{3m^2_i-3\Delta^2-m^2_j}{m^2_j\,\Delta}\,\Aij\,\right]\, , \\
\nonumber
J_3\!\!&=&\!\!\int \frac{w^2(r)}{4r^2} =
\sum_{i\,j}\frac{d_{i\,} d_j}{32\,\pi}\,\,
\frac{3\,(m^2_i+m^2_j+\Delta^2)^2-4m^2_i\,m^2_j}{4m^2_i\,m^2_j\,\Delta}
\,\,\Aij\, .
\end{eqnarray}

For integrals $B_{1,2}$ and $C_{1,2\,}$, defined by Eqs.~(\ref{ss7})
and (\ref{ss8}), one obtains the relations
\begin{eqnarray}
\nonumber
B_1\!+\!3C_1\!\!&=&\!\!\int f_1(r)f_2(r)
=\frac{3}{2}J_2\!+\!3J_3\, ,{~~~}
B_1\!+\!C_1\!=\!\int f_1(r)f_2(r) z^2 =K_1+\!K_2\, , \\
\nonumber
B_2\!+\!3C_2\!\!&=&\!\!\int f^2_2(r)=9J_3\, ,{~~~~}
B_2\!+\!C_2\!=\!\int f^2(r) z^2=3K_2\, , {~~~~}z\!=\!(\bfn\bfr)/r\, ,
\end{eqnarray}
$$
K_1 =\int \frac{3 u(r) w(r)}{2\sqrt{2}\,r^2}\,z^2\, , {~~~~}
K_2 =\int \frac{3 w^2(r)}{4\,r^2}\,z^2\, ,
$$
where $J_{2,3}$ are given in Eqs.~(A.3).  So, we have
$$
B_1=\frac{3}{2}\left(K_1+\!K_2-\frac{1}{2}\,J_2-\!J_3\right) ,
{~~}
C_1=\frac{1}{2}\left(\frac{3}{2}\,J_2+\!3J_3-\!K_1-\!K_2\right) ,
$$
$$
B_2=\frac{9}{2}\,(K_2-\!J_3) , {~~~~}C_2=\frac{3}{2}\,(3J_3-\!K_2) ,
\eqno {\rm (A.4)}
$$

\vspace{-5mm}
\begin{eqnarray}
\nonumber
K_1\!\!&=&\!\sum_{i\,j}\frac{3\,c_{i\,}d_j}{16\sqrt{2}\,\pi}\,
\left[\,\frac{-m^2_i}{2m_j \Delta^2} +\frac{m_i}{m^2_j}\,\Lij -
\frac{(m^2_i\!-\!m^2_j)^2+3\Delta^4}{2m^2_j\Delta^3}\,\Aij\,\right],\\
\nonumber
K_2\!\!&=&\!\sum_{i\,j}\frac{3\,d_{i\,} d_j}{32\,\pi}\,\left[
\frac{-(m^2_i\!+\!m^2)(m^2_i\!-\!m^2_j)^2+3(m^2_i\!+\!m^2_j)\Delta^4
+2\Delta^6}{4m^2_i\,m^2_j\,\Delta^3}\,\right] \Aij\,\, .
\end{eqnarray}

\section{The functions $h_{11}(r)$, $h_{12}(r)$, $h_1(r)$, and
         $h_2(r)$, Eqs.~(\ref{dd6})}
\label{sec:func}

Fourier inversion of Eqs.~(\ref{dd6}), where form factor
$F(s)$, defined in Eqs.~(\ref{a7}), gives

\vspace{-5mm}
\begin{eqnarray}
\nonumber
h_{11}(r) \!\!&=&\!\! \frac{\ear}{r}\, , {~~~~~~}
h_{12}(r) = i\Lambda^2_0\,
\frac{(1\!+\!\Lambda r)\eLr\!-\!(1\!-\!\alpha r)\ear}{r^3\,(\LLaa)}\, ,
{~~~~~~~~~~~~~~~~~~~~~~~~~~~~~~~~~~}(A.5) \\
\nonumber  & & \\
\nonumber
h_1(r) &=&\!\!\Lambda^4_0\left[\frac{(1+\Lambda r)\,\eLr}
{2r^3\,(\LLaa)}\!+\!
\frac{h_0(-\!\Lambda,r)\!-\!h_0(\alpha,r)}{(\LLaa)^2}\right] , {~~~}
h_0(\alpha,r)=\frac{3\!-\!3\alpha r\!+\!\alpha^2 r^2}{r^5}\,\ear\, , \\
\nonumber  & & \\
\nonumber
h_2(r) &=&\!\!-\Lambda^4_0 \left[\frac{\eLr}{2r^3\,(\LLaa)} +
\frac{(1\!+\!\Lambda r)\eLr\!-\!(1\!-\!\alpha r)\ear}
{r^5\,(\LLaa)^2}\right] , {~~~~}
\alpha=\left\{\!\begin{array}{rl}
-a_h & \!(a^2_h\!>\!0) \\ \vspace{-4mm} \\ i |a_h| & \!(a^2_h\!<\! 0)
\end{array}\!\right. .
\end{eqnarray}

\section{The integrals $A_{11\,}$, $A_{12\,}$, $A_{221\,}$,
         $A_{222\,}$, and $A_{223\,}$, Eqs~(\ref{dd7})}
\label{sec:int2}

Using Eqs.~(A.1), (A.5), and (A.2), one obtains the expressions
for integrals~(\ref{dd7}).  Separately, we express them for two
cases of parameter $\alpha$, used in Eqs.~(A.5), \textit{i.e.},
$\alpha\!=\!-a$ ($a^2_h\!>\!0$) and $\alpha\!=\!ia$
($a^2_h\!<\!0$), where $a\!=\!|a_h|$.

\vspace{1mm}
1) {\underline {$A_{11}\,$.}} For the integral $A_{11}$, we
obtain
$$
A_{11\,}(-a)=\scijp\,
\left[\,-\frac{2 (m_{ij}\!+\!a)}{Q}\,A_0(Q,a) -L_0(Q,a)\,\right] ,
{~~~}\eqno{(A.6)}
$$ $$
A_{11\,}(\,ia)=\scijp\,\left[\,a_1(Q)+a_1(-Q)+ib_1(Q)+ib_1(-Q)\right],
$$ $$
a_1(x)=-\,\frac{(a\!+\!x)L(x)\!+\!m_{ij} A(x)}{x}\, , {~~~~}
b_1(x)=\,\frac{(a\!+\!x)A(x)\!-\!m_{ij} L(x)}{x}\, .
$$
Hereafter in this Appendix, we use the notation
$$
A_0(x,y)={\rm arctg}\,\frac{x}{m_{ij}\!+\!y}\, ,{~~~~}
L_0(x,y)=\ln \left[(m_{ij}\!+\!y)^2\!+\!x^2\right]\, , \eqno{(A.7)}
$$ $$
A(x)={\rm arctg}\,\frac{a\!+\!x}{m_{ij}}\, , {~~~~~~}
L(x)=\frac{1}{2}\ln \left[m^2_{ij}+(a\!+\!x)^2\right]\, .
$$

\vspace{1mm}
2) {\underline {$A_{12}\,$.}}  The integral $A_{12}$ can be
written as
$$
A_{12\,}(\alpha)=\LLLa\,
[J_{12\,}(-\Lambda)\!-\!J_{12\,}(\alpha)]\, , {~~~~}
J_{12\,}(\alpha)=i\int\!f^2(r)\,\frac{(1\!-\!\alpha r)}{r^3}\,\ear
\eqno{(A.8)}
$$
(in this Appendix, we use the notation from Eqs.~(\ref{dd7}) and
Eqs.~(\ref{dd8})), and
$$
J_{12}(-a)=\scijp
\left[\left(\frac{2a^3\!-\!3a^2 m_{ij}\!+\!m^3_{ij}}{3Q^2}\!+\!
m_{ij}\right) A_0(Q,a) +\frac{Q}{3} L_0(Q,a)\,\right] , \eqno{(A.9)}
$$ $$
J_{12}(\,ia)=\scijp
\left[\, a_1(Q) +a_1(-Q) +ib_1(Q) +ib_1(-Q)\right] ,
$$ $$
a_1(x)=u(x)\,L(x)\!+\!v(x)\,A(x)\, , {~~~~}
b_1(x)=v(x)\,L(x)\!-\!u(x)\,A(x)\, ,
$$ $$
u(x)\!=\!\frac{a^3}{3x^2}\!+\!\frac{x}{3}\, , {~~~~} v(x)\!=\!
\left(\frac{3 a^2\!+\!m^2_{ij}}{3x^2}\!+\!1\right)\frac{m_{ij}}{2}.
$$

\vspace{1mm}
3) {\underline {$A_{221\,}$ and $A_{222\,}$.}} The integrals
$A_{221}$ and $A_{222}$ can be written as
$$
A_{221\,}(\alpha)\!=\!\frac{1}{2}\,(3\, Z_2\!-\!Z_1)\, , {~~~~~}
A_{222\,}(\alpha)\!=\!\frac{1}{2}\,(Z_1\!-\!Z_2)\, , \eqno{(A.10)}
$$ $$
Z_1=\int\!f^2(r)\,r^2 h_{1\,}(r)=\!\LLLL\left[\,Z_{10}+
\frac{Z_{11\,}(-\Lambda)\!-\!Z_{11\,}(\alpha)}{\LLaa}\,\right] ,
$$ $$
Z_{10}=\!\int\!f^2(r)\frac{1\!+\!\Lambda r}{2 r}\,\eLr\, ,
{~~~~~} Z_{11\,}(\alpha)=\!\int\!f^2(r)\,r^{2\,}h_{0\,}(r)\, ,
$$ $$
Z_2=\int\!f^2(r)\,(\bfm\bfr)^{2\,} h_{1\,}(r)=\!\LLLL\left[\,Z_{20}
+\frac{Z_{21\,}(-\Lambda)\!-\!Z_{21\,}(\alpha)}{\LLaa}\,\right] ,
$$ $$
Z_{20}=\!\int\!f^2(r)\,(\bfm\bfr)^2\frac{1\!+\!\Lambda r}{2r^3}\,\eLr
\, ,{~~~~} Z_{21\,}(\alpha)=\!\int\!f^2(r)\,(\bfm\bfr)^2\,h_{0\,}(r)
$$
(see $h_{0\,}(r)$ and $h_{1\,}(r)$ in Eqs.~(A.5)). For integrals
$Z_{10}$ and $Z_{11}$ one gets
$$
Z_{10}\!=\!-\scijp \left[\,\frac{1}{2}\,L_0(Q,\Lambda)\!
+\!\frac{m_{ij}}{Q}\,A_0(Q,\Lambda)\,\right] , \eqno{(A.11)}
$$

\vspace{-4mm}
\begin{eqnarray}
\nonumber
Z_{11\,}(-a)\!\!\!&=&\!\!\!\scijp
\left[\,\frac{Q^2\!+\!a^2\!-\!3m_{ij}}{2}\,L_0(Q)\!+\!
\frac{3Q^2\!+\!a^2\!-\!m_{ij}}{Q}\,A_0(Q)\,\right] , \\
\nonumber
Z_{11\,}(\,\,ia)\!\!&=&\!\!\scijp\,
\left[\, a_1(Q) +a_1(-Q) +ib_1(Q) +ib_1(-Q)\right] ,
\end{eqnarray}
$$
a_1(x)=u(x)\,L(x)\!+\!v(x)\,A(x)\, , {~~~~~}
b_1(x)=v(x)\,L(x)\!-\!u(x)\,A(x)\, ,
$$ $$
u(x)=\frac{1}{2}\,(x^2\!-\!a^2\!-\!3\,m_{ij})\, , {~~~~~}
v(x)=\frac{m_{ij}}{2Q}\,(3x^2\!-\!a^2\!-\!m^2_{ij})\, .
$$
For integrals $Z_{20}$ and $Z_{21}$, we have
$$
Z_{20}\!=\!\scijp\left[-\frac{2m_i m_j}{3Q^2}-\frac{1}{6}
L_0(Q,\Lambda)\!+\!\frac{(m_{ij}\!-\!2\Lambda)(\mL)^2}{3Q^3}
A_0(Q,\Lambda)\right] , \eqno{(A.12)}
$$

\vspace{-4mm}
\begin{eqnarray}
\nonumber
Z_{21\,}(-a)\!\!\!&=&\!\!\!\scijp\Biggl[\,\frac{m_i m_j}{Q^2}
\left(\frac{7a^2}{15}+\frac{3m_{ij\,} a}{10}\right) +
\left(\frac{3 Q^2}{10}+\frac{a^2\!-\!3 m_{ij}}{6}\right)\, L_0(Q) \\
\nonumber
 &+&\!\!\!\left(\frac{3m_{ij}}{2} Q +\frac{m_{ij}\,(3m^4_{ij}
 -10m^2_{ij} a^2+15a^4)+8a^5}{30\,Q^3}\right) A_0(Q)\Biggr] , \\
\nonumber
Z_{21}(\,\,ia)\!\!&=&\!\!\!\scijp\,\Biggl[\,\frac{m_i m_j}{Q^2}
\left( -\frac{7\,a^2}{15}-i\frac{3 m_{ij} a}{10}\right)
+ a_1(Q) +a_1(-Q) +ib_1(Q) +ib_1(-Q)\Biggr] ,
\end{eqnarray}
$$
a_1(x)=u(x)\,L(x)\!+\!v(x)\,A(x)\, , {~~~~~}
b_1(x)=v(x)\,L(x)\!-\!u(x)\,A(x)\, ,
$$ $$
u(x)=\frac{3 x^2}{10}\!-\frac{a^2\!+\!3\,m^2_{ij}}{6}\!+\!
\frac{2a^5}{15\,x^3}\, , {~~}
v(x)=\frac{3m_{ij}}{4} x\!+\!\frac{m_{ij}}{60\,x}\,
[3 m^4_{ij}\!+\!10 m^2_{ij} a^2\!+\!15 a^4] .
$$

\vspace{1mm}
4) {\underline {$A_{223}$.}} The integral $A_{223}$ can be
written as
$$
A_{223}(\alpha)\!=\int\!f^2(r)\,r^2\,h_{2\,}(r)=\!\LLLL\,
\left[ J_{30}\!+\!\frac{J_{31}(-\Lambda)\!-\!J_{31}(\alpha)}{\LLaa}
\right]\, , \eqno{(A.13)}
$$ $$
J_{30}=\int\!f^2(r)\,\frac{\eLr}{2r}\, , {~~~~~} J_{31\,}(\alpha)
=\int\!f^2(r)\,\frac{1\!-\!\alpha r}{r^3}\,\ear
$$
(see $h_{2\,}(r)$ in Eqs.~(A.5)). The expressions for $J_{30}$ and
$J_{31}$ have the forms
$$
J_{30}=\frac{1}{2}\scijp\,\left[\,-\frac{2 (m_{ij}\!+\!\Lambda)}{Q}\,
A_0(Q,\Lambda) -L_0(Q,\Lambda)\,\right] , \eqno{(A.14)}
$$

\vspace{-4mm}
\begin{eqnarray}
\nonumber
J_{31}(-a)\!\!\!&=&\!\!\!\scijp
\left[\left(\frac{Q^2}{6}\!+\!\frac{a^2\!-\!m^2_{ij}}{2}\right)
L_0(Q,a)+\!\left(m_{ij\,}Q\!+\!\frac{a^3\!+\!3\,m_{ij\,} a^2\!-\!
m^3_{ij}}{3\,Q}\right) A_0(Q,a) \right] , \\
\nonumber
J_{31}(\,\,ia)\!\!&=&\!\!\scijp\,
\left[\,a_1(Q)+a_1(-Q)+ib_1(Q)+ib_1(-Q)\right] ,
\end{eqnarray}
$$
a_1(x)=u(x)\,L(x) + v(x)\,A(x)\, , {~~~~}
b_1(x)=v(x)\,L(x) - u(x)\,A(x)\, ,
$$ $$
u(x)=\frac{x^2}{6}-\frac{a^2\!+\!m^2_{ij}}{2}-\frac{a^3}{3x}\, , {~~~}
v(x)=\frac{m_{ij\,}x}{2} -\frac{3\,m_{ij\,}a^2\!+\!m^3_{ij}}{6x}\, .
$$

\section{The integral $A_{11\,}$ with NKE terms}
\label{sec:int3}

\begin{figure}[th]
\centering{
\includegraphics[height=0.2\textwidth, angle=0]{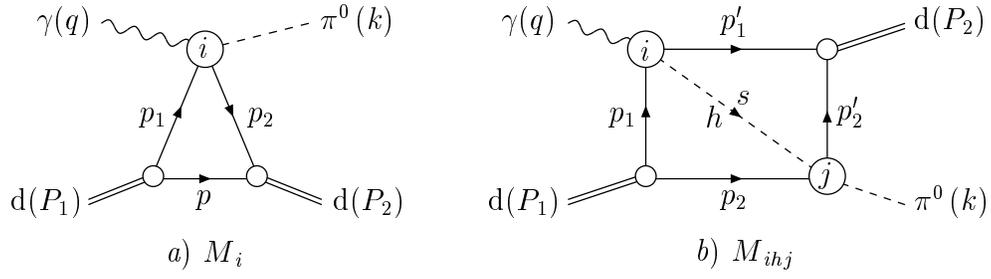}
}\caption{Feynman diagrams for the \gdpid\ reaction.
          (a) single- and (b) double-scattering.
          \label{fig:g1}}
\end{figure}

The integral $A_{11}$ can be written in $p$-space as
$$
A_{11}=\!\int\!\ddbfp\,\frac{\varphi(q_1)\,\varphi(q_2)}{2\,D}\, ,
{~~}
D\!=\!(\bfp'_1+\!\bfp_2)^2\!+\!\gamma\,(\bfp^2_1+\!\bfp^2_2)\!
+\!a^2_h\!-\!i0\, , {~~} \gamma\!=\!\frac{Q_0}{m}\,  \eqno{(A.15)}
$$
(see $a^2_h$ and $Q_0$ in Eqs.~(\ref{dd4})), where $\bfp'_1\,$,
$\bfp_2$ are CM 3-momenta of intermediate nucleons in
Fig.~\ref{fig:g1}b, $\bfq_1\!=\!\bfp_2\!+\!\half\bfq$,
$\,\bfq_2\!=\!-\bfp'_1\!-\!\half\bfk$ and $\varphi(q_{1,2})$
are $s$ DWF's.  We use the Gaussian DWF given in $p$- and
$r$-space by the relations:
$$
\varphi(p)\!=\!A_0\,\exp(-ap^2)\, ,{~~~}\psi(r)\!=\!B_0\,\exp(-br^2)\, ,
{~~~} b\!=\!\frac{1}{4a}\, , \eqno{(A.16)}
$$ $$
N_S\!=\!\intdp\,\varphi^2(p) =\!\int\!d^3\bfr\,\psi^2(r)=
\frac{A^2_0}{(8\pi a)^{3/2}}=B^2_0\left(\frac{\pi}{2b}\right)^{3/2}
=\!\scij{4\pi\, (m_i\!+\!m_j)}\, ,
$$
where $N_S$ is normalization of $s$-wave part of the Bonn DWF,
and $c_i$ and $m_i$ are the parameters from Eqs.~(A.1). For
denominator
$D$ in Eqs.~(A.15), we can use representation
$$
\frac{1}{D}=i\!\int\limits^{\infty}_0\!e^{-{\dist iDt}} dt\, .
$$
Then, calculating the Gaussian integrals $\int\!d^3\bfp'_1
d^3\bfp_2\,$, we finally obtain
$$
A_{11}=\int\limits^{\infty}_0\! dt\,J(t)\, ,{~~~~}
J(t)=\frac{i\,A^2_0}{2\,(4\pi)^3\,D^{3/2}_0}\,
\exp\left[ -\frac{ia\,t\,C_0}{4D_0}-ia^2_{h\,}t\,\right] ,
\eqno{(A.17)}
$$ $$
D_0=a^2\!-\!\gamma (2\!+\!\gamma)\,t^2 + 2ia\,(1\!+\!\gamma)\,t\, ,
{~~~}
C_0=4a\,Q^2+\gamma\left[\,a\,(q^2\!+\!k^2)+
it\,(2\!+\!\gamma)\,k^2\right] ,
$$
%
where $Q\!=\!\half |\bfq\!+\!\bfk|$.  When $a^2_h\!=\!0$
(``static" approximation) and $Q\!=\!0$, one can express
$A_{11}$ through $<\!1/r\!>_d$.  Writing $A_{11}$ in $r$-space,
we obtain
$$
A_{11}(a^2_h\!=\!Q\!=\!0)=\half\intra\,\psi^2(r)=
\frac{N_S}{8\pi}\left<\!\frac{1}{r}\!\right>_d\, . \eqno{(A.18)}
$$
We use the Bonn DWF in order to obtain the value $<\!1/r\!>_d$
and get the slope parameter $b$ of the effective Gaussian
DWF~(A.16) through the relations:
$$
\left<\!\frac{1}{r}\!\right>_d=-\scij{4\pi}\,\ln (m_i\!+\!m_j)\, ,
{~~~~} b=\frac{\pi}{8}\left<\!\frac{1}{r}\!\right>^2_d\, .
\eqno{(A.19)}
$$


\newpage
\newpage
\begin{figure}[th]
\centering{
\includegraphics[height=0.5\textwidth, angle=90]{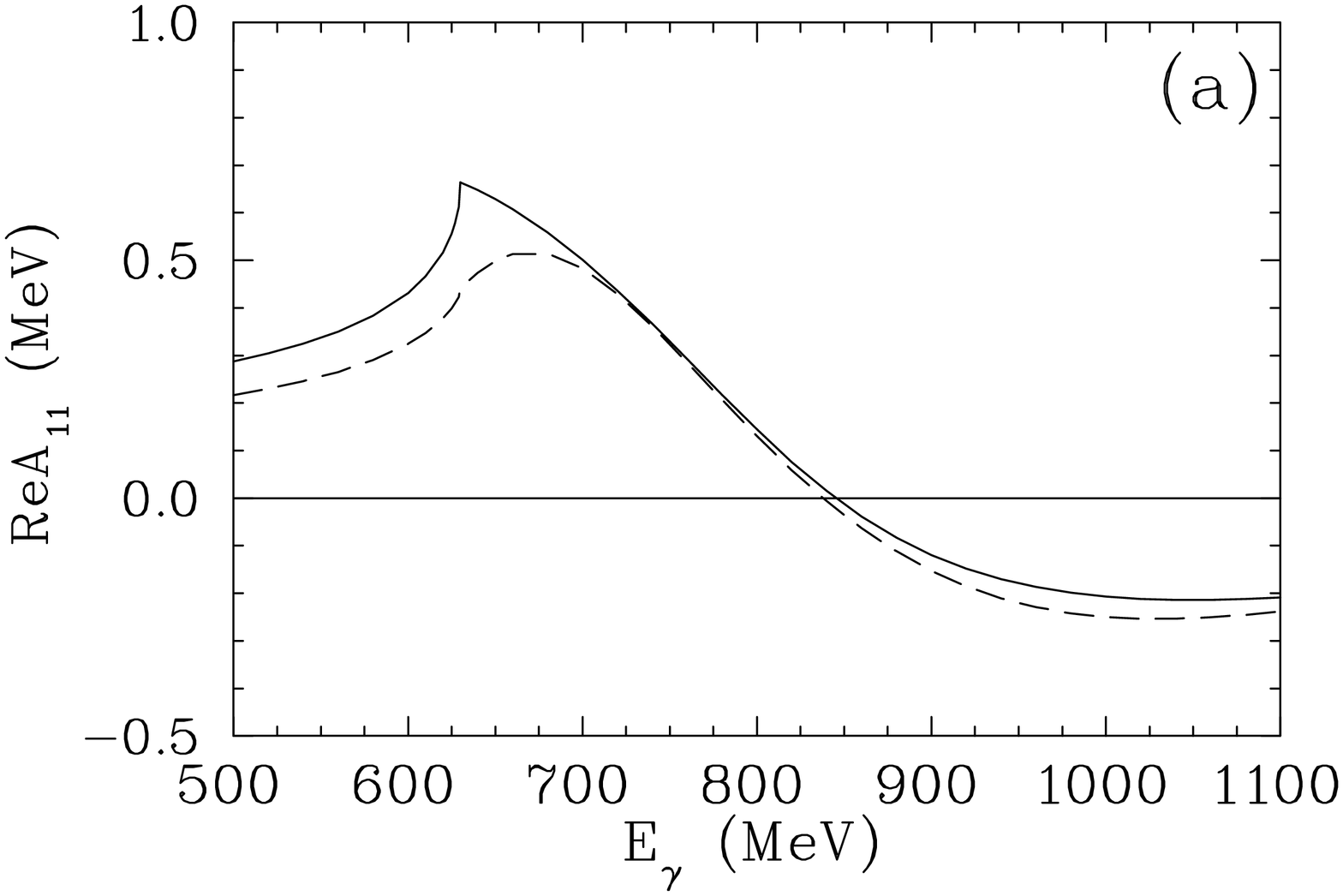}\hfill
\includegraphics[height=0.5\textwidth, angle=90]{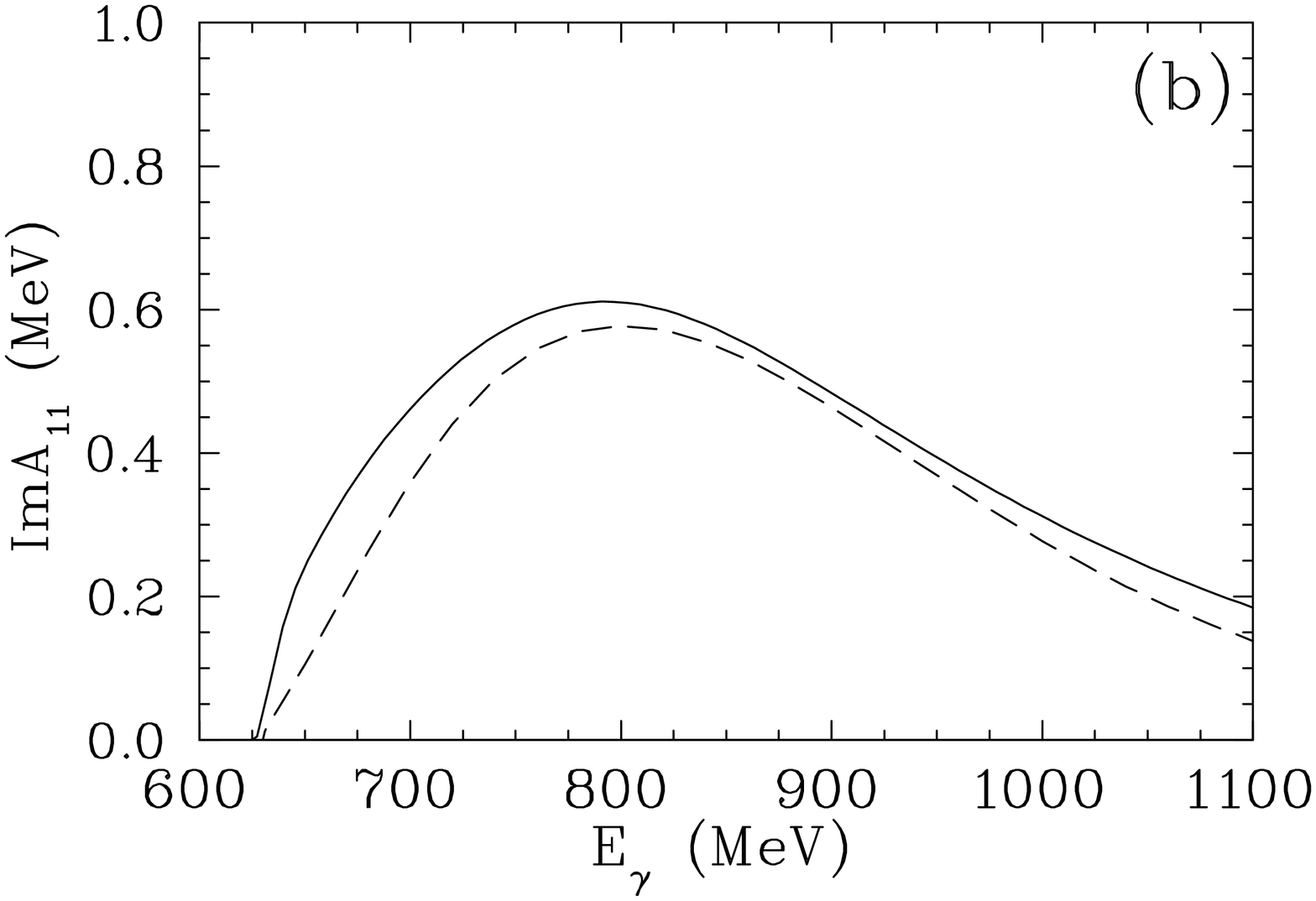}
\includegraphics[height=0.5\textwidth, angle=90]{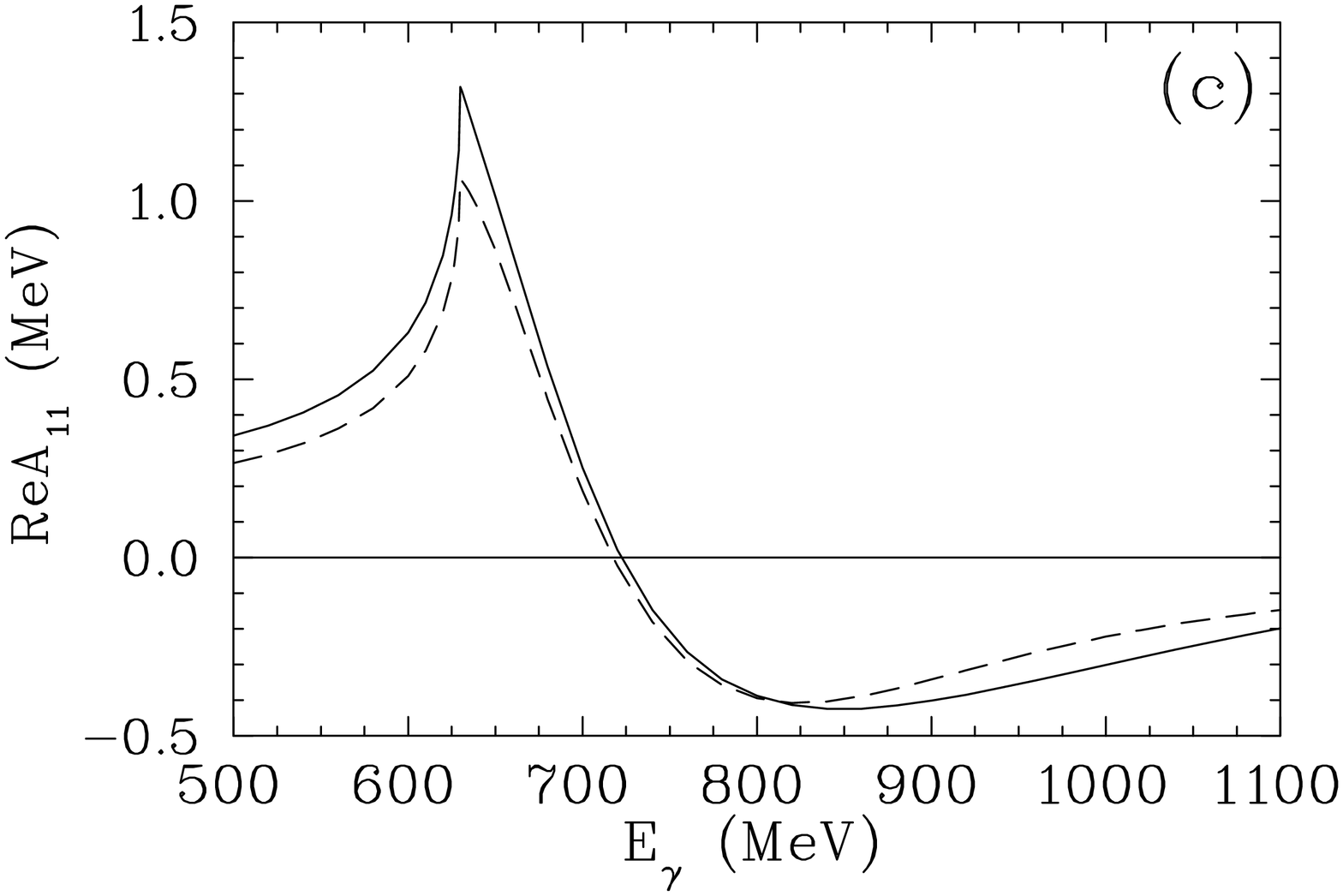}\hfill
\includegraphics[height=0.5\textwidth, angle=90]{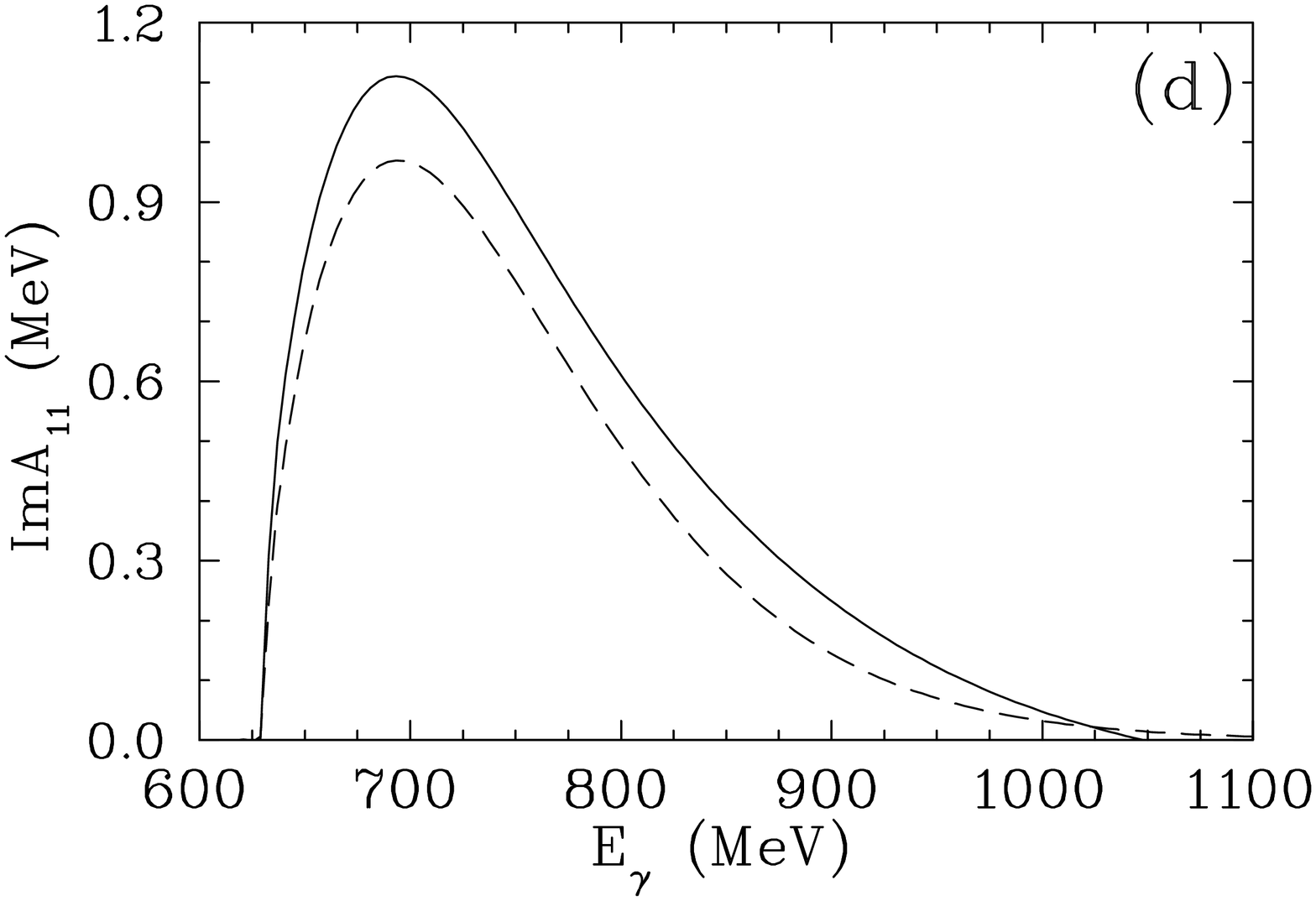}
\includegraphics[height=0.5\textwidth, angle=90]{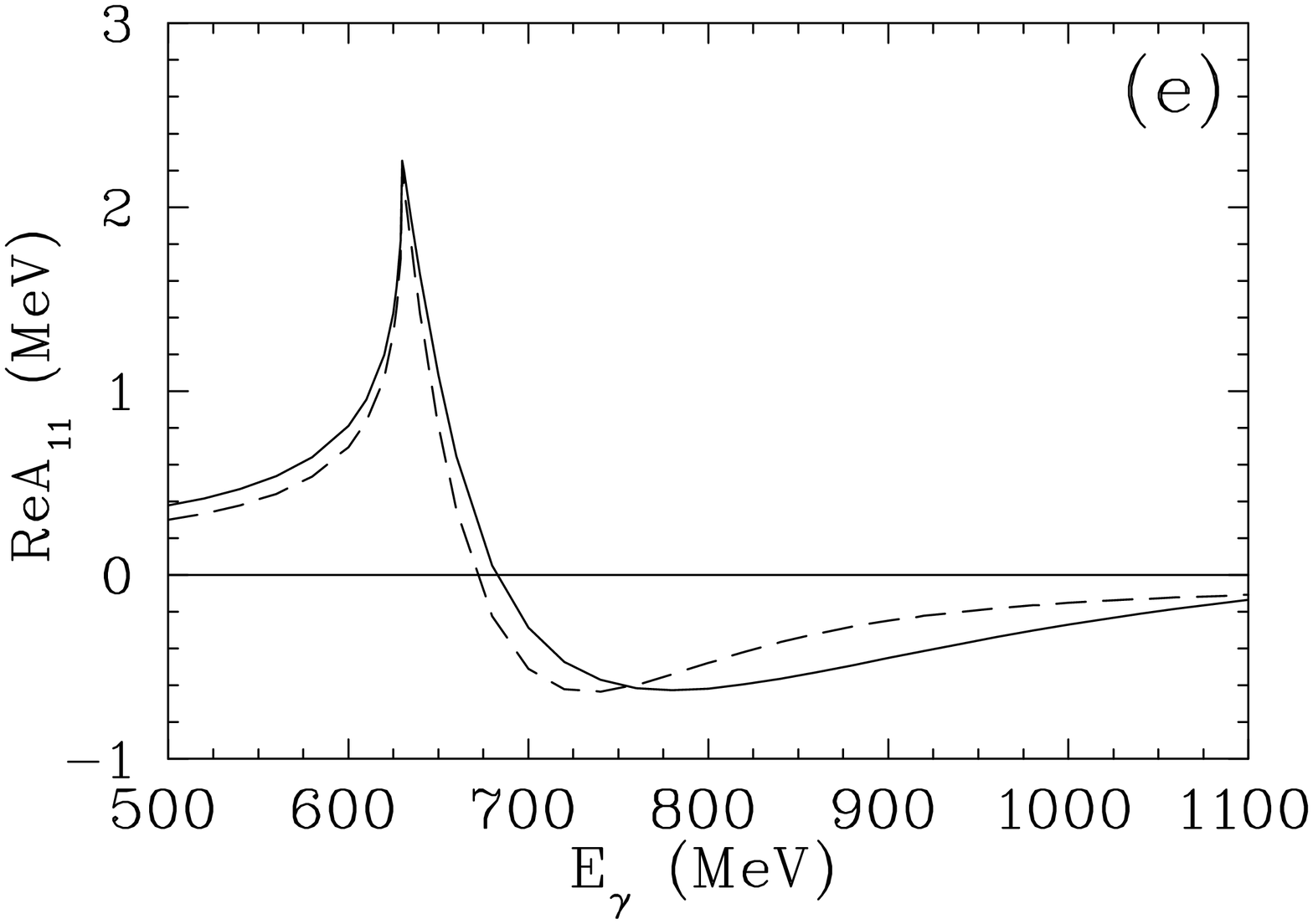}\hfill
\includegraphics[height=0.5\textwidth, angle=90]{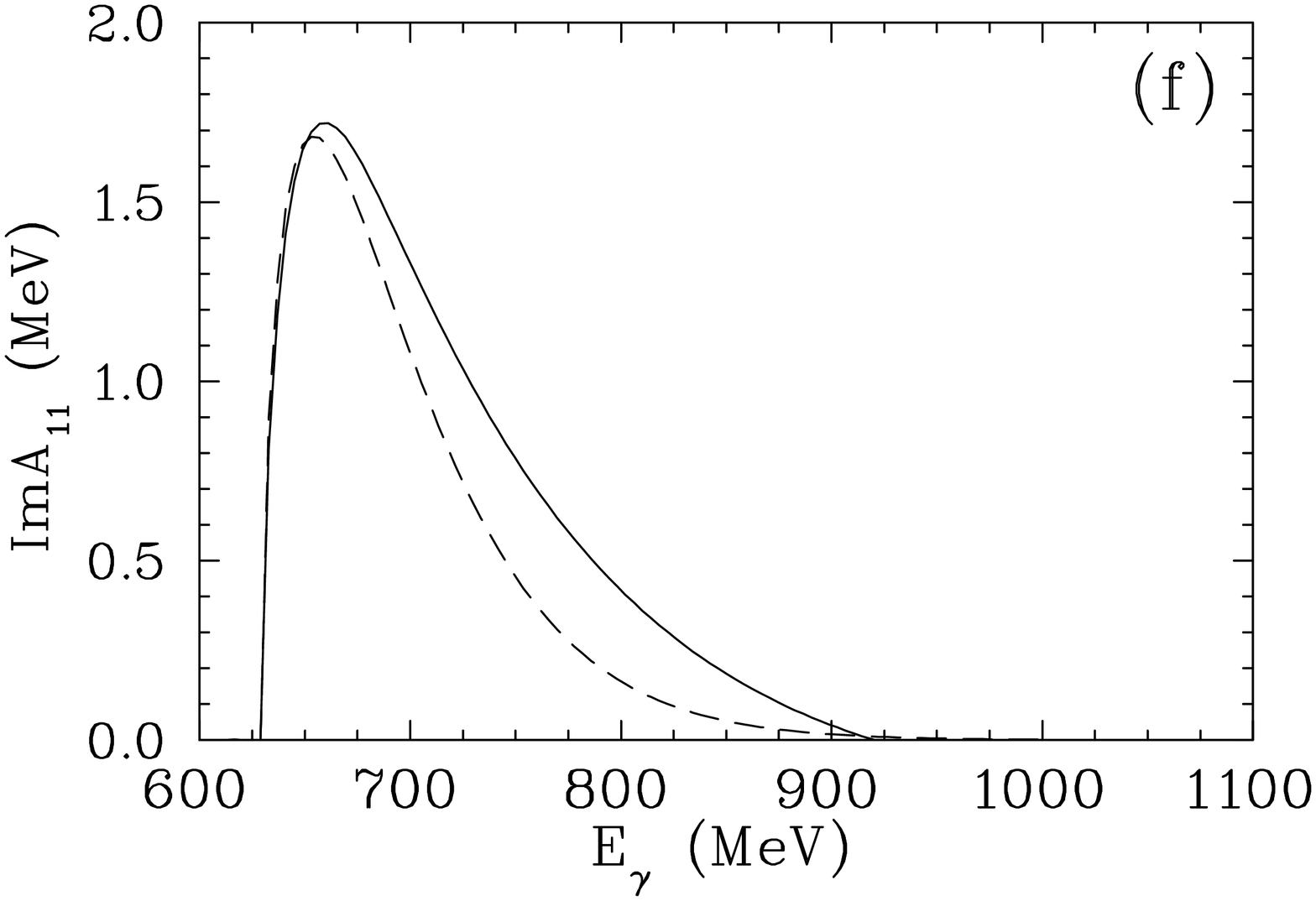}
}\caption{(a), (c), and (e) [(b), (d), and (f)] show real
          [image] parts of the loop integral $A_{11}$
          of the double-scattering diagram (Fig.~
          \protect\ref{fig:g1}b) in ``static"
          approximation with intermediate $\eta$ meson
          in \reac.  The results are given for the
          values (a) and (b) $z\!=\!\cos\theta=\!0\,$;
          (c) and (d) $z=-0.55$; and (e) and (f)
          $z=-0.85$.  Solid (dashed) curves correspond
          to the results with the Bonn (Gaussian) $s$ DWF.
          \label{fig:g4}}
\end{figure}
\newpage
\newpage
\newpage


\begin{thebibliography}{99}

\bibitem{CLAS} Y.~Ilieva {\it et al.} [CLAS Collaboration],
               to be published in Nucl.\ Phys.\
               \textbf{A} (2004) [nucl-ex/0309017].
\bibitem{Walker}R.~L.~Walker, Phys.\ Rev.\ \textbf{182}, 1729 (1969).
\bibitem{Moor} R.~G.~Moorhouse, H.~Oberlack, and A.~H.~Rosenfeld,
               Phys.\ Rev.\ D\ \textbf{9}, 1 (1974).
\bibitem{Garc} H.~Garcilazo and E.~Moya~de~Guerra, Nucl.\ Phys.\
               \textbf{A562}, 521 (1993).
\bibitem{Drech}D.~Drechsel, O.~Hanstein, S.~S.~Kamalov, and
               L.~Tiator, Nucl.\ Phys.\ \textbf{A645}, 145 (1999)
               [nucl-th/9807001].
\bibitem{Garc1}H.~Garcilazo and E.~Moya~de~Guerra, Phys.\ Rev.\
               C\ \textbf{52}, 49 (1995).
\bibitem{Kamal}S.~S.~Kamalov, L.~Tiator, and C.~Bennhold, Phys.\ Rev.\
               C\ \textbf{55}, 98 (1997) [nucl-th/9602023].
\bibitem{Knoch}G.~Kn\"ochlein, D.~Drechsel, and L.~Tiator,
               Z.\ Phys.\ A\ \textbf{352}, 327 (1995) 327
               [nucl-th/9506029].
\bibitem{Benme}M.~Benmerrouche, N.~C.~Mukhopadhyay, and J.~F.~Zhang,
               Phys.\ Rev.\ D\ \textbf{51}, 3227 (1995)
               [hep-ph/9412248].
\bibitem{Ritz} F.~Ritz and H.~Arenh\"ovel, Phys.\ Lett.\
               \textbf{B447}, 15 (1999) [nucl-th/9810027]; \\
               Phys.\ Rev.\ C\ \textbf{64}, 034005 (2001)
               [nucl-th/0011089].
\bibitem{Benn} C.~Bennhold and H.~Tanabe, Nucl.\ Phys.\ \textbf{A530},
               625 (1991).
\bibitem{Kond} L.~A.~Kondratyuk and F.~M.~Lev, Yad.\ Fiz.\ \textbf{23},
               1056 (1976) [Phys.\ At.\ Nucl. (former Sov.\ J.\
               Nucl.\ Phys.) \textbf{23}, 556 (1976)]; \\
               Yad.\ Fiz.\ \textbf{27}, 831 (1978) [Phys.\ At.\
               Nucl. (former Sov.\ J.\ Nucl.\ Phys.) \textbf{27},
               441 (1976)].\\
               L.~A.~Kondratyuk, F.~M.~Lev, and L.~V.~Shevchenko,
               Yad.\ Fiz.\ \textbf{36}, 377 (1982) [Phys.\ At.\
               Nucl. (former Sov.\ J.\ Nucl.\ Phys.) \textbf{36},
               220 (1982)].
\bibitem{pide} B.~M.~Abramov \textit{et al.}, Nucl.\ Phys.\
               \textbf{A372}, 301 (1981).\\
               R.~Keller \textit{et al.}, Phys.\ Rev.\ D\
               \textbf{11}, 2389 (1975).\\
               M.~Akemoto \textit{et al.}, Phys.\ Rev.\ Lett.\
               \textbf{50}, 400 (1983).
\bibitem{Iman} A.~Imanishi \textit{et al.}, Phys.\ Rev.\ Lett.\
               \textbf{54}, 2497 (1985).
\bibitem{PDG}  S.~Eidelman \textit{et al.} [Particle Data
               Group], Phys.\ Lett.\ \textbf{B592}, 1
               (2004); \hbox{http://pdg.lbl.gov}.
\bibitem{sm02}  R.~A.~Arndt, W.~J.~Briscoe, I.~I.~Strakovsky,
                and R.~L.~Workman, Phys.\ Rev.\ C\ \textbf{66},
                055213 (2002) [nucl-th/0205067].
\bibitem{Tar}  V.~E.~Tarasov, V.~V.~Baru, and A.~E.~Kudryavtsev,
               Yad.\ Fiz.\ \textbf{63}, 871 (2000) [Phys.\
               Atom.\ Nucl.\ \textbf{63}, 801 (2000)].
\bibitem{Bonn} R.~Machleidt \textit{et al.}, Phys.\ Rep.\
               \textbf{149}, 1 (1987).
\bibitem{Tar1} This ``$c$-transformation" is applied to nucleon spin
               and isospin operators when writing the expression for
               the amplitude, we move along the nucleon lines
               (anticlockwise here) of the diagram in the direction
               of the nucleon momentum~\protect\cite{Tar}.
\bibitem{Arndt}R.~A.~Arndt, R.~L.~Workman, Z.~Li, and L.~D.~Roper,
                Phys.\ Rev.\ C\ \textbf{42}, 1864 (1990).
\bibitem{CR}   R.~L.~Crowford and W.~T.~Morton, Nucl.\ Phys.\
               \textbf{B211}, 1 (1983).
\bibitem{David}R.~Davidson, N.~C.~Mukhopadhyay, and R.~Wittman,
               Phys.\ Rev.\ D\ \textbf{43}, 71 (1991).
\bibitem{Dumbra}O.~Dumbrajs \textit{et al.}, Nucl.\ Phys.\
               \textbf{B216}, 277 (1983).

\end{thebibliography}
\end{document}